\documentclass[
superscriptaddress,
reprint,
amsmath,amssymb,
aps,
pra
]{revtex4-2}

\usepackage[normalem]{ulem}
\usepackage{hyperref,xcolor}
\usepackage{color}

\usepackage{physics}
\usepackage{graphicx}
\usepackage{setspace}
\graphicspath{ {./Figures/} }

\usepackage{dcolumn}
\usepackage{bm}
\usepackage{amsthm}

\usepackage{tikz}
\usetikzlibrary{arrows.meta,bending,decorations.pathreplacing} 
\tikzset{>={Stealth[length=3pt,width=4pt]}}
\usetikzlibrary{decorations.pathreplacing}
\usepackage{empheq,mathtools}

\usepackage{caption}
\usepackage{subcaption}
\usepackage{adjustbox}
\usepackage{pgfplots}
\usepackage{amsmath}
\usepackage{amssymb}
\pgfplotsset{compat=1.18}
\usepackage{ragged2e}

\usepackage{bbm}

\newcommand{\id}{{\mathbbm 1}}

\newtheorem{definition}{Definition}
\newtheorem{theorem}{Theorem}
\newtheorem{lemma}[theorem]{Lemma}
\newtheorem{proposition}{Proposition}
\newtheorem{corollary}{Corollary}[proposition]
\usepackage[normalem]{ulem}
\usepackage{hyperref,xcolor}
\usepackage{color}
\usepackage{mathrsfs}

\usepackage{physics}
\usepackage{graphicx}
\usepackage{setspace}
\graphicspath{ {./Figures/} }

\usepackage{dcolumn}
\usepackage{bm}
\usepackage{amsthm}
\usepackage{tikz}
\usepackage{empheq,mathtools}

\usepackage{caption}
\usepackage{subcaption}
\usepackage{adjustbox}
\usepackage{pgfplots}
\usepackage{amsmath}
\usepackage{amssymb}
\pgfplotsset{compat=1.18}
\usepackage{ragged2e}

\usepackage{bbm}

\newcommand{\PS}{\mathfrak{P}} 
\newcommand{\StateSpace}{\Omega} 
\newcommand{\VecSpace}{\mathcal{V}} 

\begin{document}
\title{Non-unique decompositions of mixed states and deterministic energy transfers}

\author{Zihan Wang}
\affiliation{Department of Physics, City University of Hong Kong, Tat Chee Avenue, Kowloon, Hong Kong SAR}

\author{Fei Meng}
\email{fei.meng@glasgow.ac.uk}
\affiliation{School of Physics and Astronomy, University of Glasgow, Glasgow G12 8QQ, United Kingdom}
\affiliation{Department of Physics, City University of Hong Kong, Tat Chee Avenue, Kowloon, Hong Kong SAR}

\author{Oscar Dahlsten}
\email{oscar.dahlsten@cityu.edu.hk}
\affiliation{Department of Physics, City University of Hong Kong, Tat Chee Avenue, Kowloon, Hong Kong SAR}
\affiliation{Institute of Nanoscience and Applications, Southern University of Science and Technology, Shenzhen 518055, China}

\begin{abstract}
We investigate the impact of non-unique decompositions of mixed states on energy transfer. Mixed states generally have non-unique decompositions into pure states in quantum theory and, by definition, in other non-classical probabilistic theories. We consider energy transfers constituting deterministic energy harvesting, wherein the source transfers energy to the harvester but not entropy. We use the possibility of non-unique decompositions to derive that if source states in a set jointly lead to deterministic energy harvesting for the given harvesting system and interaction, then that set can be expanded to include both mixtures and superpositions of the original states in the set. As a paradigmatic example, we model the source as an EM mode transferring energy to a 2-level system harvester via the Jaynes-Cummings model. We show that the set of coherent EM mode states with fixed $|\alpha|$ that jointly achieve deterministic energy transfer can be expanded to include all mixtures and superpositions of those states. More generally, the results link the defining feature of a non-classical probability theory with the ability to achieve energy transfer without entropy transfer. 
\end{abstract}

\maketitle

\noindent {\em \bf Introduction.---}Non-unique decompositions of mixed states are a defining feature of non-classical probabilistic theories with importance in quantum information science~\cite{hardy2001quantum,plavala2023general,jiang2024unification,jiang2024framework,plavala2022operational,lin2020necessity,barnum2015post}. For example, in quantum theory, $|c_0|^2\ket{0}\bra{0}+|c_1|^2\ket{1}\bra{1}=\frac{1}{2}(c_0\ket{0}+c_1\ket{1})(c_0^*\bra{0}+c_1^*\bra{1})+\frac{1}{2}(c_0\ket{0}-c_1\ket{1})(c_0^*\bra{0}-c_1^*\bra{1})$ for any pure states denoted $\ket{0}$ and $\ket{1}$. More generally, we say there is a non-unique decomposition of mixed state $\rho$ when $\rho=\sum_i p_i\mu_i=\sum_i p'_i\mu'_i$ where the pure states $\mu$ in the decompositions are not equal: $\{\mu'_i\}\neq \{\mu_i\}$. Non-unique decomposition is exploited in quantum information science protocols, including Wiesner's quantum money~\cite{wiesner1983conjugate} and BB84 quantum key distribution~\cite{bennett2014quantum}, to hide information from forgers/eavesdroppers. Non-unique decomposition is furthermore the defining feature of a non-classical probability theory in the generalised probabilistic theories framework~\cite{hardy2001quantum,plavala2023general,jiang2024unification,jiang2024framework}.

Energy transfer between physical systems is a central theme across physical sciences, including in thermodynamics and in energy harvesting~\cite{pan2024mechanical,ullah2022review,wei2017comprehensive,liu2019intelligently,siebert2005self,meng2025quantum,skrzypczyk2014work,nielsen2010quantum,cohen2019quantum}. Of particular relevance here is the case of energy harvesting, where the aim is to absorb energy from some source for use in a load system that stores the energy or uses it for work. A key challenge in energy harvesting is the randomness and variability of many energy sources~\cite{braak2016semi,boukobza2005entropy,castano2022entropy,swati2022quantifying,Mitcheson2008,wei2017comprehensive}. Loads typically require well-defined input, such as predictable voltage. The standard mitigation for such sources involves rectifiers, which in regimes of low voltage lead to significant losses~\cite{horowitz1989art,Szarka2012,surender2021rectenna}. Those losses are, in part, fundamental in that without them the rectifiers could generate DC voltage from thermal fluctuations, in violation of the second law of thermodynamics~\cite{liu2019intelligently}. A qualitatively different approach to condition the output of such sources, applicable to certain classes of non-thermal sources, involves using a quantum system as an intermediary~\cite{meng2025quantum}. The multitude of paths between ground and excited state of the quantum system can be exploited such that a multitude of source behaviours all achieve transfer from the ground state to the excited state at the same time, harvesting energy deterministically, without entropy, from a source with entropy. That deterministic harvesting approach does not have fundamental losses, unlike the rectifiers or Maxwell's-demon-like active interventions~\cite{meng2025quantum,bennett1982thermodynamics}. A key question concerning the deterministic energy harvesting approach is for what classes of sources can the approach be employed?

We here tackle that question by exploiting the indistinguishability of decompositions of source states and convexity arguments. We argue that (i) if a set of source states jointly lead to deterministic harvesting, any probabilistic (i.e.\ convex) mixture of them must also do so, (ii) any pure state in any decomposition of those mixed states must then also do so. The argument applies to generalised Hamiltonian mechanics of general probabilistic theories~\cite{hardy2001quantum,plavala2023general,jiang2024unification,jiang2024framework,plavala2022operational,lin2020necessity,barnum2015post}, including quantum theory. In the case of quantum theory, the argument implies that any superposition of states that individually lead to deterministic energy harvesting also leads to the same deterministic harvesting. In that sense, theories with non-unique decomposition of source states have a comparative abundance of source states that jointly lead to deterministic harvesting. We extend the argument to approximately indistinguishable decompositions, showing that the conclusions are robust under perturbations of source states in the quantum case. As a paradigmatic example, we model the source and harvester in the widely applicable Jaynes-Cummings model, fully quantising the discussion in Ref.~\cite{meng2025quantum}, showing that coherent states $\ket{\alpha}$ on the source with the same $|\alpha|$ and their superpositions like $\ket{\alpha}+\ket{-\alpha}$ jointly lead to deterministic harvesting.

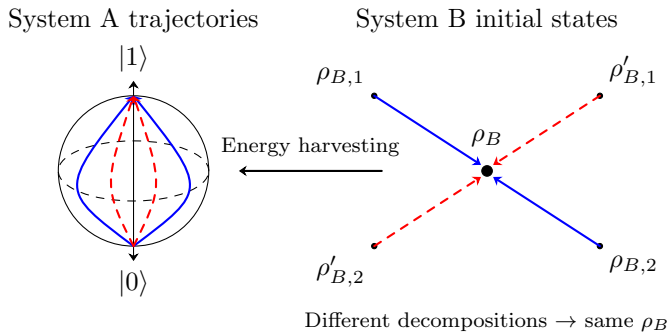
\begin{figure}[htbp!]

\centering

\begin{tikzpicture}[line cap=round, line join=round]

\begin{scope}
  \coordinate (Acenter1) at (0,0);
  \draw (Acenter1) circle (1);
  \draw[dashed] (Acenter1) ellipse (1 and 0.4);
  \draw[->] (Acenter1) -- ++(0,1.2) node[above] {$|1\rangle$};
  \draw[->] (Acenter1) -- ++(0,-1.2) node[below] {$|0\rangle$};

  \draw[thick, blue, ->] (0,-1) .. controls (-1,-0.25) .. (0,1);
  \draw[thick, blue, ->] (0,-1) .. controls (1,-0.25) .. (0,1);
  \draw[thick, red, dashed, ->] (0,-1) .. controls (-0.4,-0.2) .. (0,1);
  \draw[thick, red, dashed, ->] (0,-1) .. controls (0.4,-0.2) .. (0,1);


  \node at (0,2) {System A trajectories};
\end{scope}

\begin{scope}[xshift=4.7cm]
  \coordinate (center) at (0,0);
  
  \filldraw (center) circle (2pt) node[above,yshift=6pt] {$\rho_B$};
  
  \coordinate (A1) at (-1.5,1);
  \coordinate (A2) at (-1.5,-1);
  \filldraw (A1) circle (1pt) node[above left] {$\rho_{B,1}$};
  \filldraw (A2) circle (1pt) node[below left] {$\rho_{B,2}'$};
  \draw[->, thick, blue] (A1) -- (-0.08,0.08);
  \draw[->, thick, red, dashed] (A2) -- (-0.08,-0.08);

  \coordinate (B1) at (1.5,1);
  \coordinate (B2) at (1.5,-1);
  \filldraw (B1) circle (1pt) node[above right] {$\rho_{B,1}'$};
  \filldraw (B2) circle (1pt) node[below right] {$\rho_{B,2}$};
  \draw[->, thick, red, dashed] (B1) -- (0.08,0.08);
  \draw[->, thick, blue] (B2) -- (0.08,-0.08);
  
  \node at (0,2) {System B initial states};
  \node at (0,-2) {\footnotesize Different decompositions $\to$ same $\rho_B$};
\end{scope}

\draw[->, thick] (3.3,0) -- (1.4,0) node[midway,above,yshift=2pt] {\footnotesize Energy harvesting};

\end{tikzpicture}

\caption{\justifying
{\bf Non-unique decomposition and deterministic energy transfer.}
Illustration of the relationship between different decompositions of a fixed mixed state \(\rho_B\) in system~B and the induced trajectories in system~A under the energy harvesting protocol. The two decompositions— \(\{\rho_{B,1}, \rho_{B,2}\}\) represented by solid blue lines and \(\{\rho_{B,1}', \rho_{B,2}'\}\) represented by dashed red lines—correspond to different pure-state ensembles that realize the same \(\rho_B\). Each pure state in a decomposition drives a different trajectory of the qubit in system~A at a designated time, shown in the matching colors.}

\label{fig:1}
\end{figure}


\noindent {\em \bf Generalized probabilistic theories (GPTs).---}GPTs provide a unifying operational framework encompassing classical mechanics, quantum theory, and hypothetical post-quantum models. This general structure enables the study of energy transfer processes without restricting to the Hilbert space formalism. In GPTs for mechanical systems with phase space variables \( z = (q, p) \), such as classical or quantum phase space dynamics, states can be taken to be normalized real-valued functions \( W(z) \), and effects (the generalisation of POVM elements) as functionals assigning outcome probabilities via inner products \( \int W(z) g(z) \, dz \)~\cite{jiang2024unification,jiang2024framework,henselder2005star,li2013weyl}. Mixtures, i.e.\ convex combinations of states, are by definition allowed. General dynamics are represented by affine transformations of the states. The generalised Hamiltonian evolution equation reads
\begin{equation}
\frac{\partial W}{\partial t} = \int K(k) W \sin\left( \frac{k}{2} \Lambda \right) H \, dk,
\label{eq:generalhamiltonianevolution}
\end{equation}
where \( \Lambda \) is the Poisson bracket operator, $H(q,p)$ the Hamiltonian function, and \( K(k) \) is a theory-specific kernel that reduces to \( (2/k)\delta(k - \hbar) \) in quantum theory \cite{jiang2024unification,jiang2024framework,henselder2005star,li2013weyl}. Energy eigenstates are defined as the most pure stationary states—extreme points of the set of time-invariant distributions—which coincide with quantum eigenstates in the quantum case~\cite{jiang2024unification,jiang2024framework}.

An important example of a state is the Wigner function
\(
W(q,p) \coloneqq \frac{1}{2\pi \hbar} \int dx\, e^{i p x/\hbar} \langle q - \tfrac{1}{2} x | \hat{\rho} | q + \tfrac{1}{2} x \rangle,
\)
which represents the density matrix \(\hat{\rho}\) in phase space, embedding quantum theory into the phase-space GPT framework.

This phase-space GPT formulation allows us to analyze how structural features of the theory—particularly non-unique decompositions of source states—affect the energy transfer in the harvesting.

\noindent{\em \bf Deterministic energy harvesting (DEH).---}In previous work, a paradigmatic example of DEH via quantum systems was analyzed in a semiclassical regime, where the harvester is a qubit and the source is treated as a classical drive~\cite{meng2025quantum}. In the paradigmatic example based on Rabi oscillation, the energy transfer is governed by the semiclassical Hamiltonian (under the rotating wave approximation and resonance)~\cite{meng2025quantum}:
\begin{equation}\label{eq:semiclassicalH}
H_C^{(\phi)} = \frac{\hbar \omega_0}{2} \hat{\sigma}_z + 2 A \cos(\omega t + \phi)\,\hat{\sigma}_x ,
\end{equation}
which generates an evolution driving deterministic excitation of the qubit. The trajectories depend on the source phase $\phi$, but all lead to the excited state of the qubit at the same time.  This is in contrast with linear oscillators, which pick up the source phase and fail to achieve non-trivial DEH~\cite{meng2025quantum}.

In the present work, we extend this regime to include the source and to a largely theory-independent setting. We consider a bipartite system with subsystems $A$ (harvester) and $B$ (source), each described by a normalized quasiprobability distribution over phase space. The initial state of $A$ is fixed as $W_A^{(0)}(q_A, p_A)$, while $B$ is prepared in one of several possible distributions $\{ W_B^{(i)}(q_B, p_B) \}$. The joint system evolves under a fixed Hamiltonian evolution of $AB$ represented in phase space by $\Phi_\tau$, corresponding to a Hamiltonian $H = H_A + H_B + H_{\text{int}}$, which acts reversibly on the composite distribution according to Eq.(\ref{eq:generalhamiltonianevolution}). By definition, a set of source states on $B$ jointly achieve deterministic energy harvesting to $A$ exactly if 
$A$ is changed from a pure lower energy state to a pure higher energy state at the time $\tau$ for any source state in the set.

\begin{definition}[DEH in the phase-space formalism]
\label{def:DEH_Wigner}
Given an initial product state
\[
W_{AB}^{(i)}(q_A, p_A; q_B, p_B)
= W_A^{(0)}(q_A, p_A) \cdot W_B^{(i)}(q_B, p_B),
\]
and the phase space distribution propagator \(\Phi_\tau\) for evolution time \(\tau\), we say DEH is achieved if
\begin{align}
\label{eq:DEHachieved}
\int dq_B\, dp_B\, 
\Phi_\tau[W_{AB}^{(i)}(q_A, p_A; q_B, p_B)]
= W_A^{(1)}(q_A, p_A), \quad \forall i,
\end{align}
where \(W_A^{(1)}\) is the excited-state distribution of \(A\).  
This ensures that \(A\) is driven deterministically from \(W_A^{(0)}\) to \(W_A^{(1)}\) for all \(W_B^{(i)}\) in the designated set.
\end{definition}


\noindent{\em \bf Non-unique decomposition of mixed states in the source gives deterministic energy harvesting.---}A central conceptual question concerns the \emph{role of non-unique decomposition} of the source state. In the phase-space formulation, a mixed source $W_B(q_B, p_B)$ can often be expressed as distinct convex combinations of pure-state phase-space distributions:
\begin{equation}
\label{eq:decompositions}
W_B = \sum_i p_i W_B^{(i)} = \sum_j q_j \widetilde{W}_B^{(j)},
\end{equation}
where $\{W_B^{(i)}\}$ and $\{\widetilde{W}_B^{(j)}\}$ are non-identical sets of pure phase-space distributions. 

However, in a DEH-compatible evolution, if each $W_B^{(i)}$ individually drives subsystem $A$ from $W_A^{(0)}$ to $W_A^{(1)}$, then \emph{any} convex combination of these sources yields the same final $W_A^{(1)}$, by linearity of the phase-space evolution, as depicted in Fig.~\ref{fig:1}. We formalize this as follows.

\begin{proposition}[Decomposition-Irrelevance of DEH]
\label{prop:DEH_GPT_Decomposition}
Let $\{W_B^{(i)}\}$ be a set of pure source states that jointly, for the same dynamics $\Phi_\tau$, achieve DEH (Def.~\ref{def:DEH_Wigner}).
Then DEH is achieved for (i) any convex combination $W_B = \sum_i p_i W_B^{(i)}$, and
(ii) any pure states appearing in equivalent decompositions of 
$W_B$, and their convex combinations. 
\end{proposition}


\noindent\textit{Sketch of proof:} 
For (i), the linearity of the dynamics ensures that a convex combination of pure source states achieving DEH yields the same convex combination of identical outcomes, and thus also achieves DEH. For (ii), if the same mixed state is expressed in an alternative decomposition (Eq.~\ref{eq:decompositions}), linearity again implies that the final marginal is a convex combination of the outcomes of the new extremal states. Since this marginal is independent of the decomposition, each extremal state must individually yield the same outcome. Therefore, all pure states in any alternative decomposition, together with their mixtures, achieve DEH (Details in SM).

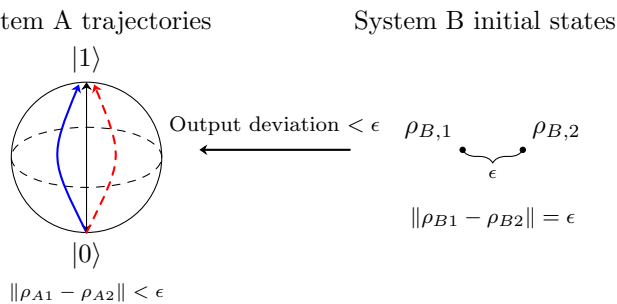
\begin{figure}[htbp!]

\centering

\begin{tikzpicture}[line cap=round, line join=round]

\begin{scope}[yshift=-4.5cm]
  \coordinate (Acenter2) at (0,0);
  \draw (Acenter2) circle (1);
  \draw[dashed] (Acenter2) ellipse (1 and 0.4);
  \draw[->] (Acenter2) -- ++(0,1) node[above] {$|1\rangle$};
  \draw[->] (Acenter2) -- ++(0,-1) node[below] {$|0\rangle$};
  \draw[thick, blue, ->] (0,-1) .. controls (-0.5,0) .. (-0.1,0.97);
  \draw[thick, red, dashed, ->] (0,-1) .. controls (0.5,0) .. (0.1,0.97);
  \node at (0,-1.8) {\scriptsize $D(\rho_{A1},\rho_{A2}) \leq \epsilon$};
  \node at (0,1.8) {System A trajectories};
\end{scope}

\begin{scope}[xshift=5cm,yshift=-4.5cm]
  \coordinate (B1) at (0,0.1);
  \coordinate (B2) at (0.8,0.1);

  \filldraw (B1) circle (1pt) node[above left] {$\rho_{B,1}$};
  \filldraw (B2) circle (1pt) node[above right] {$\rho_{B,2}$};

  \draw [decorate,decoration={brace,amplitude=5pt,mirror}] (B1) -- (B2) 
    node[midway,below,yshift=-4pt] {\scriptsize $\epsilon$};

  \node at (0.3,1.8) {System B initial states};

  \node at (0.4,-0.8) {\footnotesize $D(\rho_{B1}, \rho_{B2}) = \epsilon$};
\end{scope}

\draw[->, thick] (3.5,-4.4) -- (1.5,-4.4) node[midway,above,yshift=2pt] {\footnotesize Output deviation $< \epsilon$};

\end{tikzpicture}

\caption{\justifying {\bf Robustness under source state perturbation for the quantum case}. When two input states to system B differ by a small amount (i.e., $D(\rho_{B1}, \rho_{B2}) = \epsilon$), the corresponding output states \(\rho_{A1}, \rho_{A2}\) in system A will not differ more: \(D(\rho_{A1}, \rho_{A2})\leq \epsilon\)}.

\label{fig:2}
\end{figure}


\noindent{\em \bf Robustness of DEH under Source Perturbation.---}An important operational question concerns the sensitivity of DEH to small deviations in the source state.

\begin{proposition}[Distinguishability of initial source states bounds distinguishability of final harvester states]
\label{prop:DEH_GPT_Perturbation}
Consider an initial harvester state $W_A(0)$ and an independently prepared initial source state that is either
$W_B^{(1)}(0)$ or $W_B^{(2)}(0)$. Let the joint AB system then undergo a generalised Hamiltonian evolution until time $\tau$, a map $\Phi_\tau$. Let the corresponding final marginal state on A be denoted as $W_A^{(1)}(\tau)$ or $W_A^{(2)}(\tau)$ respectively. Then
\begin{equation}\label{eq:DEH_D_bound}
\mathrm{D}(W_A^{(1)}(\tau),W_A^{(2)}(\tau))\leq \mathrm{D}(W_B^{(1)}(0),W_B^{(2)}(0)),
\end{equation}
where $\mathrm{D(.,.)}$ denotes the statistical distance that generalises the trace distance: $\mathrm{D}(W, W') \coloneqq \sup_{\{h_i\}} \sum_i \left| P(h_i|W) - P(h_i|W') \right|$, the maximal difference in probabilities over possible measurements.\\
\noindent Sketch of proof: The generalized Hamiltonian evolution $\Phi_\tau$ preserves $\int W_{AB}^{(1)}(z)\,W_{AB}^{(2)}(z)\,dz$~\cite{jiang2024framework}, where $z$ represents the AB phase space variables, implying, one can show, that $\mathrm{D}(W_{AB}^{(1)},\,W_{AB}^{(2)})$ is invariant. Discarding subsystem $B$ (integrating over $q_B,p_B$) cannot increase $\mathrm{D}$. Combining these two facts gives the stated bound (details in SM).
\end{proposition}

Eq.~(\ref{eq:DEH_D_bound}) has different implications in the quantum and classical cases because of differences in $\mathrm{D}(W_B^{(1)}(0),W_B^{(2)}(0))$. Two distinct pure classical states are fully distinguishable with $\mathrm{D}(W_B^{(1)}(0),W_B^{(2)}(0))=1$, the maximal value. Then Eq.~(\ref{eq:DEH_D_bound}) does not constrain $\mathrm{D}(W_A^{(1)}(\tau),W_A^{(2)}(\tau))$, consistent with the known possibility of strong initial state dependence (classical chaos). In the quantum case two pure states are in general not fully distinguishable and $\mathrm{D}(W_B^{(1)}(0),W_B^{(2)}(0))$ can be small. Then Eq.(~\ref{eq:DEH_D_bound}) is a significant constraint, giving a guarantee of low sensitivity to perturbations of the initial source state as depicted in Fig.~\ref{fig:2} (details in SM). A qualitatively similar argument can be made in the quantum case for the relative entropy as a distinguishability measure (details in SM).



\noindent{\em \bf DEH in density matrix quantum theory.---}We now reformulate DEH using density matrices, representing the key quantum case of the GPT framework, in particular   Def.~\ref{def:DEH_Wigner} of DEH and Proposition~\ref{prop:DEH_GPT_Decomposition} of the decomposition-irrelevance of DEH.

While the semiclassical model already exemplifies the phase-independent energy extraction described in the GPT framework, it does not capture genuinely quantum effects such as entanglement, coherence, and non-unique decompositions of mixed states. Our present approach therefore extends DEH to a fully quantized setting, with $U = e^{-i H_Q \tau}$ and $H_Q = H_A + H_B + H_{\rm int}$ representing the total Hamiltonian.

Let the initial state be 
$\rho_A (0) \otimes \rho_B^{(i)}$, with $\rho_A (0) = |g\rangle\langle g|$ and $\rho_B^{(i)}$ a possibly mixed source state.  
\begin{definition}[DEH in Density Matrix Formalism]
\label{def:DEH_quantum}
Let $\{\rho_B^{(i)}\}$ be a set of source states. DEH is achieved at $t=\tau$ if
\[
\rho_A(\tau) = \Tr_B \big[ U(\rho_A \otimes \rho_B^{(i)}) U^\dagger \big] = |e\rangle\langle e|, \quad \forall i,
\]
where $U = e^{-i H_Q \tau}$ is the total unitary evolution. The process succeeds for each $\rho_B^{(i)}$.
\end{definition}

\begin{proposition}[Convex Closure of DEH States]
\label{prop:DEH_quantum_convexclosure}
Let Def.~\ref{def:DEH_quantum} hold. If DEH holds for a set $\{\rho_B^{(i)}\}$, it also holds for any convex combination
$\rho_B = \sum_i p_i \rho_B^{(i)}$, as well as for any alternative decomposition of the same mixed state.
\end{proposition}

\noindent
(See Supplemental Material for the proof. )

Proposition \ref{prop:DEH_quantum_convexclosure}  highlights a key quantum feature and serves as the quantum version of proposition~\ref{prop:DEH_GPT_Decomposition}: a single mixed source can admit multiple valid decompositions, all supporting DEH. Such decomposition-independence has no apparent analog in semiclassical models and underscores the necessity of a fully quantum formulation to capture the full operational structure of DEH. From it, we have the following corollary:

\begin{corollary}[Superposition-generated alternative DEH ensembles]
\label{cor:DEH_alternative_ensembles}
If DEH holds for a set of pure states $\{|\phi_B^{(i)}\rangle\}$, then it also holds for any superpositions ($\,\sum_ic_i|\phi_B^{(i)}\rangle$ where $c_i\in \mathbb{{C}}$) of these states.
\end{corollary}

\noindent
\textit{Sketch of proof:} For two states, complementary superpositions such as $|\phi_B^{(1)}\rangle \pm |\phi_B^{(2)}\rangle$ cancel cross terms when equally mixed, reproducing the same density operator. The same cancellation mechanism generalizes to larger sets, where suitable constructions of many-term superpositions eliminate all cross terms. Since DEH depends only on $\rho_B$, every such alternative superposition ensemble also achieves DEH (details in SM). 

Knowledge of Corollary \ref{cor:DEH_alternative_ensembles} can significantly increase the set of source states. Mixed states $\rho$ of $d$ basis states have $d$ free parameters  ($d-1$ if normalisation is included), whereas superpositions and mixtures of the same states have $d^2$ free parameters ($d^2-1$ if normalisation is included), as can be seen from direct calculation taking into account that $\rho=\rho^{\dagger}$ (see~\cite{hardy2001quantum} for discussion).

\begin{figure}[htbp!]
    \centering
    \begin{subfigure}{0.9\linewidth}
        \centering
        \begin{tikzpicture}
\begin{axis}[
    width=0.85\linewidth,
    height=0.45\linewidth,
    xlabel={\(g t\)},
    ylabel={Fidelity},
    xmin=0, xmax=25,
    ymin=0, ymax=1.05,
    xtick distance=10,
    ytick distance=0.2,
    legend style={draw=none, at={(0.5,1.05)}, anchor=south, font=\small, legend columns=-1},
    tick label style={font=\small},
    label style={font=\small},
    domain=0:25,
    samples=2000
]

\addplot [
    only marks,
    mark=triangle*,
    mark size=4pt,
    draw=black,
    fill opacity=0,
    line width=0.5pt,
    mark repeat=10,
] 
table [x=t, y=fidelity1, col sep=space] {fidelity_data.dat};
\addlegendentry{\(\phi=0\) \quad}

\addplot [
    only marks,
    mark=*,
    mark size=2pt,
    draw=black!20!black,
    fill opacity=0,
    line width=0.5pt,
    mark repeat=5,
] 
table [x=t, y=fidelity2, col sep=space] {fidelity_data.dat};
\addlegendentry{\(\phi=\pi/4\) \quad}

\addplot [
    black,
    thick
] 
table [x=t, y=fidelity3, col sep=space] {fidelity_data.dat};
\addlegendentry{\(\phi=\pi/2\)}

\end{axis}
\end{tikzpicture}
        \caption{\justifying
        Phase-independence of DEH: same \(|\alpha|\), different \(\phi\), pure states.}
        \label{fig:DEH_pure_phase}
    \end{subfigure}

    \vspace{1em}

    \begin{subfigure}{0.9\linewidth}
        \centering
        \begin{tikzpicture}
\begin{axis}[
    width=0.85\linewidth,
    height=0.45\linewidth,
    xlabel={\(g t\)},
    ylabel={Fidelity},
    xmin=0, xmax=25,
    ymin=0, ymax=1.05,
    xtick distance=10,
    ytick distance=0.2,
    legend style={draw=none, at={(0.5,1.05)}, anchor=south, font=\small, legend columns=-1},
    tick label style={font=\small},
    label style={font=\small},
    domain=0:25,
    samples=2000
]

\addplot [black, thick, solid] table [x=t, y=fidelity1, col sep=space] {fidelity_data_2.dat};
\addlegendentry{Prep 1 \quad}

\addplot [
    only marks,
    mark=x,
    mark size=4pt,
    draw=black,
    fill opacity=0,
    line width=0.5pt,
    mark repeat=5,
] 
table [x=t, y=fidelity2, col sep=space] {fidelity_data_2.dat};
\addlegendentry{Prep 2}

\end{axis}
\end{tikzpicture}
        \caption{\justifying
        Non-unique decomposition: same \(\rho_B\), different decompositions.}
        \label{fig:DEH_decomposition}
    \end{subfigure}

    \vspace{1em}

    \begin{subfigure}{0.9\linewidth}
        \centering
        \begin{tikzpicture}
\begin{axis}[
    width=0.85\linewidth,
    height=0.45\linewidth,
    xlabel={\(g t\)},
    ylabel={Fidelity},
    xmin=0, xmax=25,
    ymin=0, ymax=1.05,
    xtick distance=10,
    ytick distance=0.2,
    legend style={draw=none, at={(0.5,1.05)}, anchor=south, font=\small, legend columns=-1},
    tick label style={font=\small},
    label style={font=\small},
    domain=0:25,
    samples=2000
]

\addplot [blue, thin, thick] table [x=t, y=fidelity1, col sep=space] {fidelity_data_3.dat};
\addlegendentry{Prep 1}

\addplot [orange, thin, thick] table [x=t, y=fidelity2, col sep=space] {fidelity_data_3.dat};
\addlegendentry{Prep 2}

\end{axis}
\end{tikzpicture}
        \caption{\justifying
        Approximate DEH under small perturbations in the field state.}
        \label{fig:DEH_approx}
    \end{subfigure}

    \captionsetup{justification=justified,singlelinecheck=false}
    \caption{\justifying
    {\bf Deterministic and approximate deterministic energy harvesting (DEH) dynamics in the Jaynes Cummings model.} Each panel displays the fidelity between the time-evolved qubit state \(\rho_A(t)\) and the excited state \(|1\rangle\), with fidelity unity corresponding to perfect energy absorption. The time axis is shown in units of \(g t\), where \(g\) is the qubit--field coupling strength.}
    \label{fig:DEH_JCM_examples}
\end{figure}
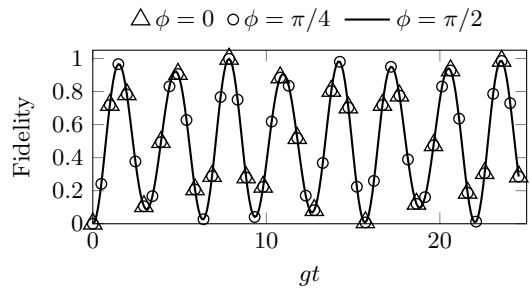
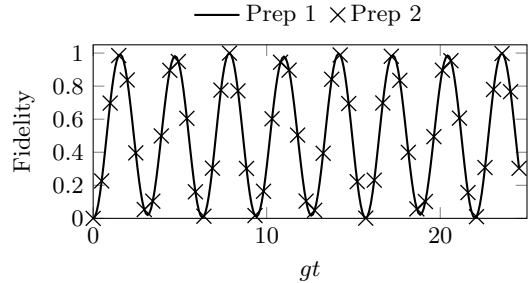
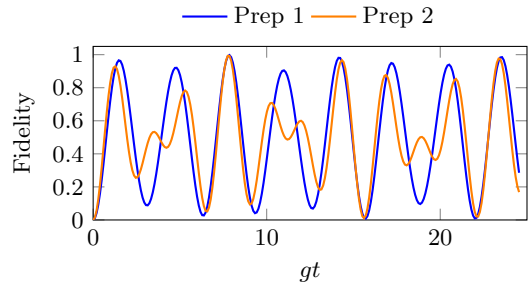

\noindent{\em \bf DEH in the Jaynes–Cummings model.---}To exemplify a fully quantized implementation of DEH, we consider the Jaynes–Cummings (JC) model , which describes a two-level system (qubit) interacting with a single quantized mode of an electromagnetic field \cite{jaynes2005comparison,phoenix1991establishment,gea1990collapse,knight2024enduring,twyeffort2022defining}. The JC Hamiltonian is
\begin{equation}
\label{eq:JC}
H_{\rm JC} = \frac{\hbar \omega_0}{2} \hat{\sigma}_z + \hbar \omega_c \hat{a}^\dagger \hat{a} + \hbar g \left( \hat{\sigma}_+ \hat{a} + \hat{\sigma}_- \hat{a}^\dagger \right),
\end{equation}
where $\omega_0$ and $\omega_c$ denote the qubit and field frequencies, and $g$ is the coupling strength.

We initialize the system in a product state $\rho_{AB}(0) = |g\rangle\langle g| \otimes \rho_B$, where $\rho_B$ is the source field state. For the example of a coherent state, $\rho_B=|\alpha\rangle\langle\alpha|$ where $|\alpha\rangle = e^{-|\alpha|^2/2} \sum_{n=0}^{\infty} \frac{\alpha^n}{\sqrt{n!}} |n\rangle$ and $n$ is the photon number. The qubit undergoes Rabi oscillations whose amplitudes depend on the photon-number distribution, yet the final state of $A$ remains insensitive to the field phase $\phi$ ($\alpha = |\alpha| e^{i\phi}$) for fixed $|\alpha|$, illustrating phase-independent energy harvesting (Fig.~\ref{fig:DEH_pure_phase}). Mixtures of coherent states with different $\phi$ thereby also achieve DEH.

The JC model illustrates how different decompositions of mixtures of source states achieving DEH also achieve DEH (Proposition~\ref{prop:DEH_quantum_convexclosure}), consistent with Proposition~\ref{prop:DEH_GPT_Decomposition}. For example, the mixture
\begin{equation}
\label{eq:rho_mixture}
\frac{1}{2} \big(\!|\alpha\rangle\langle \alpha| \!+\! |{-\alpha}\rangle\langle {-\alpha}|\! \big) \! \approx \! \frac{1}{2} \big(\! |+\rangle_\alpha \langle \!+\!|_\alpha \!+\! |-\rangle_\alpha \langle -|_\alpha \!\big)
\end{equation}

\noindent as a source state achieves DEH, where the even $|+\rangle_\alpha$ and odd $|-\rangle_\alpha$ cat states are defined as
\begin{equation}
\begin{split}
|\pm\rangle_\alpha &= \frac{1}{\sqrt{2(1 \pm e^{-2|\alpha|^2})}} \left( |\alpha\rangle \pm |-\alpha\rangle \right) \\
&\approx \frac{1}{\sqrt{2}} \left( |\alpha\rangle \pm |-\alpha\rangle \right).
\end{split}
\end{equation}
\noindent The approximation holds in the limit of large $|\alpha|$, where  $e^{-2|\alpha|^2}$ becomes negligible \cite{buvzek1992superpositions,glauber1963coherent,sudarshan1963equivalence,weedbrook2012gaussian}. For both ensemble representations in Eq.~\ref{eq:rho_mixture}, evolution under $H_{\rm JC}$ yields identical reduced states $\rho_A(t)$, consistent with the fact that the different decompositions cannot be distinguished. Wigner functions corresponding to these decompositions are visualized in the SM, highlighting the non-uniqueness of mixed-state representations. 

Beyond this theoretical perspective, the preparation and control of Schrödinger cat states and their statistical mixtures have become central themes in contemporary experimental physics~\cite{cotte2022experimental,yang2025hot,he2023fast,putterman2025hardware}.The DEH protocol thus aligns naturally with these developments, suggesting that this protocol could have concrete implications for ongoing efforts to generate and manipulate nonclassical superposition states.

The JC example also illustrates ``approximate" DEH (which is a quantum version of proposition~\ref{prop:DEH_GPT_Perturbation}): slightly different mixed source states produce qubit states that remain operationally indistinguishable (Fig.~\ref{fig:DEH_approx}).

\noindent{\em \bf Source in a thermal state.---}A particularly important family of mixed states is given by the Gibbs thermal states, \(
\rho_B^{\rm th} \;=\; \frac{1}{Z} \sum_i e^{-\beta E_i} |E_i\rangle \langle E_i|,\) where \(\quad Z = \sum_i e^{-\beta E_i},\), $\beta=\frac{1}{k_B T}$ is the inverse temperature and $\{|E_i\rangle\}$ denote the energy eigenstates of system $B$. Gibbs states cannot achieve DEH whenever the ground state $|0\rangle$ of the source cannot achieve DEH. (For the JC Hamiltonian of Eq.~\ref{eq:JC}, $\ket{g}_A\ket{0}_{B}$ is by inspection an eigenstate invariant under time evolution such that $|0\rangle_{B}$ cannot achieve DEH.) Proposition~(\ref{prop:DEH_quantum_convexclosure}) requires that all extremal states in the convex decomposition admit harvesting. Consequently, any mixture containing the ground state necessarily fails to achieve DEH. This observation reinforces the thermodynamic consistency of the DEH protocol. The Gibbs state effectively acts as a single heat reservoir, and its inability to serve as a DEH source is compatible with the Kelvin–Planck statement of the second law: no cyclic process can extract energy in the form of work exclusively from a thermal reservoir.

\noindent{\em \bf Conclusion and Outlook.---}We established a unified framework for deterministic energy harvesting (DEH) within generalized probabilistic theories (GPTs), encompassing classical, quantum, and post-quantum settings. Within this model-independent approach, we identified the structural role of non-unique decompositions of the mixtures of DEH-capable sources in enabling phase-independent, entropy-free energy transfer and proved robustness of DEH against small source perturbations in quantum theory. Extending beyond the original semiclassical protocol, we demonstrated that DEH persists in a fully quantized description, with the Jaynes--Cummings model providing a concrete example of phase-independent and decomposition-invariant energy transfer.

Our framework suggests several directions. Extensions to higher-dimensional and continuous-variable systems may uncover novel energy-transfer regimes \cite{vool2017introduction,freitas2014dynamics,blais2004cavity}, while links to resource theories of purity and entropy suppression could deepen the conceptual understanding of DEH. On the practical side, controllable platforms such as superconducting circuits and cavity QED offer routes to probe DEH experimentally.

\section*{Acknowledgements}

We are grateful to Zujin Wen, Sivapalan Chelvaniththilan, Xiangjing Liu, Shyam Dhamapurkar, Li Fei and Junhao Xu for discussions and feedback on the draft. We acknowledge support from The City University of Hong Kong (Project No. 9610623) and Research Grants Council (RGC) General Research Fund  (CityU 11300125).

\bibliographystyle{apsrev4-2}

\bibliography{reference}

\newpage
\onecolumngrid

\section*{SUPPLEMENTAL MATERIAL FOR ``NON-UNIQUE DECOMPOSITIONS OF MIXED STATES AND DETERMINISTIC ENERGY TRANSFERS''}

\appendix
\renewcommand{\thefigure}{S\arabic{figure}}
\renewcommand{\thedefinition}{S\arabic{definition}}

\setcounter{proposition}{0}
\setcounter{definition}{0}

\section{Proofs in details of the propositions in the main text}
Here, we show the detailed proof of proposition~\ref{prop:DEH_GPT_Decomposition}, proposition~\ref{prop:DEH_quantum_convexclosure} and corollary ~\ref{cor:DEH_alternative_ensembles}. The details of proposition~\ref{prop:DEH_GPT_Perturbation} will be illustrated in Appendix~\ref{sec:rigorous_distinguishability}.
\begin{proposition}[Decomposition-Irrelevance of DEH]
\label{prop:DEH_GPT_Decomposition}
Let $\{W_B^{(i)}\}$ be a set of pure source states that jointly, for the same dynamics $\Phi_\tau$, achieve DEH, meaning that
\begin{equation}
\label{eq:DEHachieved}
\int dq_B\, dp_B\, \Phi_\tau\!\left[ W_A^{(0)} W_B^{(i)} \right] 
= W_A^{(1)}, \quad \forall i.
\end{equation}
Then the same relation (Eq.~\ref{eq:DEHachieved}) is achieved for\\
(i) any convex combination 
$W_B = \sum_i p_i W_B^{(i)}$, and\\
(ii) any pure states appearing in equivalent decompositions of 
$W_B$, and their convex combinations. 
\end{proposition}

\noindent
\textit{Proof.}  
For part (i), let \( W_B = \sum_i p_i W_B^{(i)} \) be a convex combination of pure source states that individually achieve DEH. Using linearity of the dynamics \(\Phi_\tau\) and integration over phase space, we have
\[
\int dq_B\, dp_B\, \Phi_\tau[W_A^{(0)} W_B] 
= \int dq_B\, dp_B\, \Phi_\tau\Big[ W_A^{(0)} \sum_i p_i W_B^{(i)} \Big] 
= \sum_i p_i \int dq_B\, dp_B\, \Phi_\tau[W_A^{(0)} W_B^{(i)}] = W_A^{(1)}.
\]  
Hence, any convex combination of the \( W_B^{(i)} \) also achieves DEH.  

For part (ii), consider an alternative decomposition \( W_B = \sum_j q_j \widetilde{W}_B^{(j)} \) of the same mixed state. By linearity of \(\Phi_\tau\), we have
\[
W_A^{(1)} = \int dq_B\, dp_B\, \Phi_\tau[W_A^{(0)} W_B] 
= \int dq_B\, dp_B\, \Phi_\tau\Big[ W_A^{(0)} \sum_j q_j \widetilde{W}_B^{(j)} \Big] 
= \sum_j q_j \int dq_B\, dp_B\, \Phi_\tau[W_A^{(0)} \widetilde{W}_B^{(j)}].
\]  
Since the \( q_j \) are positive and sum to 1, each term must individually equal \( W_A^{(1)} \) (as we have assumed $W_A^{(1)}$ a pure state in the main text). Therefore, every pure state in any equivalent decomposition of \( W_B \) achieves DEH, and by part (i), any convex combination of these states does as well.  

\hfill\qedsymbol

\setcounter{proposition}{2}

\begin{proposition}[Convex Closure of DEH States] 
\label{prop:DEH_quantum_convexclosure}
Let \(\{\rho^{(i)}_B\}_{i=0}^{i_{\text{max}}}\) be a set of field states for which DEH holds. 
Then any convex combination
\begin{equation}
\rho_B = \sum_{i \in I} p_i \rho^{(i)}_B, \quad p_i \geq 0, \quad \sum_{i \in I} p_i = 1, \quad I \subseteq \{0,1,\ldots,i_{\max}\},
\end{equation}
also achieves DEH at \(t = \tau\). 
Moreover, any alternative decomposition of the same mixed state,
\begin{equation}
\rho_B = \sum_j q_j \tilde{\rho}^{(j)}_B,
\end{equation}
with \(\{\tilde{\rho}^{(j)}_B\}\) forming a distinct ensemble, likewise supports DEH for the qubit.
\end{proposition}

\noindent
\textit{Proof.} Let DEH hold for each state \(\rho^{(i)}_B\) in the set \(\{\rho^{(i)}_B\}_{i=0}^{i_{\max}}\). That is, the reduced qubit state at time \(t = \tau\) is \(|e\rangle\langle e|\) regardless of the choice of \(\rho^{(i)}_B\).

\noindent Now consider a convex mixture \(\rho_B = \sum_i p_i \rho^{(i)}_B\), with \(p_i \geq 0\) and \(\sum_i p_i = 1\). The initial joint state is
\begin{equation}
\rho_{AB}(0) = \rho_A \otimes \rho_B.
\end{equation}
Under unitary evolution,
\begin{equation}
\rho_{AB}(\tau) 
= U \rho_{AB}(0) U^\dagger
= U \big( \rho_A \otimes \sum_i p_i \rho^{(i)}_B \big) U^\dagger
= \sum_i p_i \, U (\rho_A \otimes \rho^{(i)}_B) U^\dagger .
\end{equation}
Taking the partial trace over subsystem \(B\),
\begin{equation}
\rho_A(\tau) 
= \Tr_B \!\left[ \rho_{AB}(\tau) \right] 
= \sum_i p_i \, \Tr_B \!\left[ U (\rho_A \otimes \rho^{(i)}_B) U^\dagger \right] 
= \sum_i p_i \, |e\rangle\langle e| 
= |e\rangle\langle e| .
\end{equation}

\noindent
\noindent Now suppose \(\rho_B = \sum_j q_j \tilde{\rho}^{(j)}_B\) is an alternative decomposition of the same mixed state. Then,
\begin{equation}
\rho_A(\tau) 
= \Tr_B \!\left[ U (\rho_A \otimes \rho_B) U^\dagger \right]
= \Tr_B \!\left[ U \big( \rho_A \otimes \sum_j q_j \tilde{\rho}^{(j)}_B \big) U^\dagger \right]
= \Tr_B \!\left[ U \big( \rho_A \otimes \sum_i p_i \rho^{(i)}_B \big) U^\dagger \right]
= |e\rangle\langle e| .
\end{equation}

\noindent Hence, the output is decomposition-independent, and DEH holds for any convex combination or ensemble realization of \(\rho_B\).
\hfill\qedsymbol

\begin{corollary}[Superposition-generated alternative DEH ensembles]
\label{cor:DEH_alternative_ensembles}
If DEH holds for a set of pure states $\{|\phi_B^{(i)}\rangle\}$, then it also holds for any superpositions of these states.
\end{corollary}

\noindent
\textit{Proof.}  
First, define
\[
|\tilde\psi^{(r)}\rangle \;=\; \sum_{k=1}^m (-1)^{r_k} a_k \, |\phi^k_B\rangle, \quad r_k \in \{0,1\}, \quad \sum_{k=1}^m |a_k|^2=1
\]
with normalization factor
\[
N_r = \langle \tilde\psi^{(r)}|\tilde\psi^{(r)}\rangle, 
\quad 
|\psi^{(r)}\rangle = N_r^{-1/2} |\tilde\psi^{(r)}\rangle.
\]
Hence,
\[
|\psi^{(r)}\rangle\langle\psi^{(s)}|
= N_r^{-1/2} N_s^{-1/2} 
\sum_{k=1}^m \sum_{l=1}^m (-1)^{r_k+s_l}a_ka_l 
\, |\phi^k_B\rangle \langle \phi^l_B|.
\]

\noindent By construction, there exists another state
\[
|\psi^{(r')}\rangle\langle\psi^{(s')}|
= N_{r'}^{-1/2} N_{s'}^{-1/2} 
\sum_{k=1}^m \sum_{l=1}^m (-1)^{r'_k+s'_l}a_k'a_l'  
\, |\phi^k_B\rangle \langle \phi^l_B|
\]
such that the cross terms 
$\sum_{k\neq l}|\phi^k_B\rangle\langle \phi^l_B|$ 
cancel when adding.  
Therefore,
\[
\frac{1}{2} \Big( 
|\psi^{(r)}\rangle\langle\psi^{(s)}| 
+ 
|\psi^{(r')}\rangle\langle\psi^{(s')}| 
\Big) 
= \sum_{k=1}^m p_k \, |\phi^k_B\rangle\langle \phi^k_B| 
= \rho_B.
\]

\noindent Since the mixed state achieves DEH, each component  
$|\psi^{(r)}\rangle\langle\psi^{(s)}|$ and  
$|\psi^{(r')}\rangle\langle\psi^{(s')}|$  
also achieves DEH separately.  

\noindent Finally, without loss of generality, the set 
$\{|\phi^k_B\rangle\}$ can be taken as a subset of the general family 
$\{|\phi^j_B\rangle\}$, so any superposition drawn from the original set likewise gives DEH.  
\hfill\qedsymbol

\section{Derivation of the distinguishability bound under source perturbations}
\label{sec:rigorous_distinguishability}

In this section, we provide a self-contained, mathematically rigorous proof of the distinguishability bound stated in Proposition~2 of the main text, i.e.\ that the deviation in the harvester's final state is bounded by the deviation of the source state, when both are measured by the operational distinguishability $D$.
We adopt the \emph{generalized phase-space} framework established in Ref.~\cite{jiang2024framework}. Our derivation proceeds from first principles, utilizing the specific definitions of states, measurements, and dynamical symmetries provided in Sec.~II and Sec.~III of Ref.~\cite{jiang2024framework}.

\subsection{The Mathematical Structure of States}

We begin by establishing the rigorous functional analytic structure of the theory.

\subsubsection{The Phase Space Domain.}
The physical system is described by a phase space manifold, denoted by $\PS$ (see Sec.~II.A of Ref.~\cite{jiang2024framework}).
 For continuous variable systems (e.g., classical mechanics or continuous-variable quantum mechanics), $\PS = \mathbb{R}^{2n}$ with canonical coordinates $z = (q_1, \dots, q_n, p_1, \dots, p_n)$.
For discrete variable systems (such as \emph{Spekkens' toy model} discussed in Sec.~IX.B of Ref.~\cite{jiang2024framework}, or standard finite-dimensional quantum mechanics), $\PS$ is a discrete set (e.g., a discrete grid $\mathbb{Z}_d \times \mathbb{Z}_d$ or a finite set of ontic states).

To maintain a unified formalism, we utilize the integral notation $\int_{\PS} d z$ to denote the sum over all phase space configurations. For discrete systems, this integral is rigorously interpreted as a summation $\sum_{z \in \PS}$.

\subsubsection{The Vector Space of Functions ($\VecSpace$).}
We define the underlying linear space exactly following the generalized probabilistic framework outlined in Sec.~II.C of Ref.~\cite{jiang2024framework}. Let $\VecSpace$ be the real vector space of all real-valued functions and generalized functions (classical pure states are represented by Dirac delta distributions) defined on $\PS$:
\begin{equation}
    \VecSpace \coloneqq \left\{ f: \PS \to \mathbb{R} \;\bigg|\; \int_{\PS} |f(z)| \, d z < \infty \right\}.
\end{equation}
The operations of vector addition and scalar multiplication are defined pointwise. For any $f, g \in \VecSpace$ and $\alpha, \beta \in \mathbb{R}$, the function $(\alpha f + \beta g)$ is defined by $(\alpha f + \beta g)(z) = \alpha f(z) + \beta g(z)$. This space contains both physical states and the differences between them.

\subsubsection{ The Set of Valid Physical States ($\StateSpace$).}
The set of physically valid states, denoted $\StateSpace$, is a convex subset of the real function space over the phase space $\PS$. Strictly following the postulates in Sec.~II.C and Sec.~III of Ref.~\cite{jiang2024framework}, a function $W(z)$ in the generalized phas-space constitutes a valid physical state if and only if it satisfies the following rigorous conditions:

\begin{enumerate}
    \item \emph{Normalization:} The total probability must be conserved (see Sec.~II.C and Sec.~III of Ref.~\cite{jiang2024framework}):
    \begin{equation}
        \int_{\PS} W(z) \, d z = 1.
    \end{equation}
    
    \item \emph{Operational Positivity:} While $W(z)$ is permitted to take negative values (as in the case of quasiprobability distributions), it must yield non-negative probabilities for the complete set of physically allowed measurements (effects) $\{e_i\}$ defined in the theory (see Sec.~II.C and Eq.~(17) of Ref.~\cite{jiang2024framework}). That is, for any valid effect $e_i$:
    \begin{equation}
        P(i|W) = e_i(W) \geq 0.
    \end{equation}
\end{enumerate}

The state space $\StateSpace$ forms a convex set (see Sec.~II.C of Ref.~\cite{jiang2024framework}): if $W_1, W_2 \in \StateSpace$, then the probabilistic mixture $p W_1 + (1-p) W_2 \in \StateSpace$ for any $p \in [0,1]$.

\subsection{ Generalized Measurements, the base norm, and the operational distinguishability}

To rigorously bound the deviation $\Delta W(z) = W_1(z) - W_2(z)$ of any two states $W_1$ and $W_2$, we must define a norm that quantifies physical distinguishability. In the generalized phase-space framework adopted by the PRA paper (Sec.~II.C), a state is operationally defined by the statistics of measurement outcomes (see Eq.~(15) in Ref.~\cite{jiang2024framework}). Therefore, the magnitude of the difference between two states is naturally quantified by the maximal difference in the probabilities they generate.

\begin{definition}[Measurement Effects]
Following Eq.~(17) of Ref.~\cite{jiang2024framework}, a measurement effect is a function $h: \PS \to \mathbb{R}$ that maps a phase space distribution to a probability. The probability $P(h|W)$ of observing the outcome associated with effect $h$, given a state $W$, is the linear functional:
\begin{equation}
    P(h|W) \coloneqq \int_{\PS} h(z) W(z) \, d z.
\end{equation}
For a function $h(z)$ to represent a physically valid detector response, it must satisfy the probability bounds for \emph{any} valid state $W \in \StateSpace$:
\begin{equation}
    0 \le \int_{\PS} h(z) W(z) \, d z \le 1.
\end{equation}
We denote the set of all physically allowed effects in the theory as $\mathcal{E}$.
\end{definition}

\begin{definition}[Complete Measurement]
A physical measurement procedure consists of a set of effects $\{h_i\}_{i \in I}$ (where $i$ indexes the possible outcomes) such that the probabilities sum to unity for any state. This implies the completeness condition (Eq.~(18) in Ref.~\cite{jiang2024framework}):
\begin{equation}
    \sum_{i \in I} P(h_i | W) = 1, \quad \forall W \in \StateSpace.
\end{equation}
By linearity, this requires that the sum of the effect functions equals the unit function on the phase space:
\begin{equation}
    \sum_{i \in I} h_i(z) = 1(z),
\end{equation}
where $1(z) = 1$ for all $z \in \PS$.
\end{definition}

\begin{definition}[Base Norm and Operational Distinguishability]
We define the norm of a vector $v \in \VecSpace$ as the maximum total statistical deviation observable by any complete measurement allowed by the theory.
\begin{equation}
    \| v \| \coloneqq \sup_{\{h_i\} \subset \mathcal{E}} \sum_i \left| \int_{\PS} h_i(z) v(z) \, d z \right|.
    \label{eq:op_norm_def}
\end{equation}
Here, the supremum is taken over all possible sets of effects $\{h_i\}$ that satisfy the completeness condition.  In the main text we write the operational distinguishability as
\begin{equation}
D(W_1,W_2) := \|W_1 - W_2\|,
\end{equation}
so that the base norm $\|W_1 - W_2\|$ defined in Eq.~\eqref{eq:op_norm_def} is exactly the statistical distance $D(W_1,W_2)$ appearing in Eq.~(5) of the main text, which generalises the trace distance (will show in the later subsections).
\end{definition}

\subsubsection{Equivalent formulations of base norm in different literature.}
The base norm defined in Eq.~\eqref{eq:op_norm_def} for the generalized-phase space is not an arbitrary choice; it is mathematically equivalent to the \emph{base norm} of the underlying ordered vector space $\VecSpace$. In the framework of GPTs, the state space $\StateSpace$ forms the base of a positive cone. As rigorously established in Propositions~3.40 and 3.42 of Ref.~\cite{plavala2023general} (see also the proof of Theorem~3.43 therein), the maximization over measurement effects (the dual norm) coincides with the base norm defined by the minimal decomposition of the vector $v$. Specifically, for a difference vector $v = W_1 - W_2$, we have the identity:
\begin{equation}
    \| v \| = \inf \left\{ \lambda + \mu \;\middle|\; v = \lambda x - \mu y, \; x,y \in \StateSpace, \; \lambda,\mu \ge 0 \right\},
\end{equation}
where the right hand side is the definition of base norm in ordered vector spaces. This demonstrates that calling Eq.(\ref{eq:op_norm_def}) base norm is well defined.

\subsubsection{ Operational distinguishability as statistical distance.}
When $v = \Delta W = W_1 - W_2$, the base norm $\| \Delta W \|$ rigorously quantifies the \emph{operational distinguishability} of the two states. It is directly related to the maximum probability $P_{\text{succ}}$ of correctly identifying the state in a single-shot discrimination protocol with equal prior probabilities:
\begin{equation}
    P_{\text{succ}} = \frac{1}{2} \left( 1 + \frac{1}{2} \| W_1 - W_2 \| \right).
\end{equation}
Thus, $\| \Delta W \| = D(W_1,W_2)$ measures the statistical advantage provided by optimal measurement strategies over random guessing. In particular, when $\| \Delta W \|=0$, the two states $W_1$ and $W_2$ cannot be discriminated in a single-shot experiment.

\subsubsection{ Relationship to Standard Distance Measures}

As demonstrated in Section~7 of the foundational work by Edwards~\cite{edwards1970operational}, the base norm on the predual of a von Neumann algebra (quantum mechanics) is the \emph{trace norm}, while for a commutative algebra (classical mechanics), it degenerates to the $L1$ norm (total variation distance) of the probability distributions. For convenience of the reader, we present the sketch of the proofs for the base norm reducing to L1 norm and trace norm in specific cases.

\paragraph{ Classical Mechanics (L1 norm).}  Consider a partition of the phase space into two regions: $\PS_+$ where the function $\Delta W(z)$ is positive, and $\PS_-$ where it is negative.
\begin{equation}
    \Delta W(z) = \Delta W_+(z) - \Delta W_-(z), \quad \text{where } \Delta W_{\pm} \ge 0.
\end{equation}
We assume the theory allows for a measurement capable of distinguishing these regions. Specifically, let us assume there exists a measurement $\{h_+, h_-\}$ such that $h_+(z) \approx 1$ on $\PS_+$ and $h_-(z) \approx 1$ on $\PS_-$. Substituting this measurement into the definition:
\begin{align}
    \sum_{i \in \{+,-\}} \left| \int_{\PS} h_i(z) \Delta W(z) \, d z \right| &= \left| \int_{\PS_+} \Delta W(z) \, d z \right| + \left| \int_{\PS_-} \Delta W(z) \, d z \right| \nonumber \\
    &= \int_{\PS_+} \Delta W_+(z) \, d z + \int_{\PS_-} \Delta W_-(z) \, d z \nonumber \\
    &= \int_{\PS} | \Delta W(z) | \, d z.
\end{align}
Therefore, the base norm is computed by integrating the absolute magnitude of the difference vector:
\begin{equation}
    \| \Delta W \|^{\text{cm}} = \int_{\PS} | W_1(z) - W_2(z) | \, d z = \|\Delta W \|_{\text{L1}}.
\end{equation}
This quantity provides a concrete measure of the \emph{distinguishability} of the states. In classical mechanics, this is the total variation distance and the base norm Eq.~\eqref{eq:op_norm_def} reduces to the L1 norm. 

\paragraph{Quantum Mechanics (Trace Distance).}
In the quantum regime, valid effects correspond to positive operators $0 \le \hat{E} \le \hat{I}$. Due to the uncertainty principle, the Wigner representation of an effect, $h(z)$, cannot be an arbitrary indicator function on the phase space (it cannot exhibit sharp features below the scale of $\hbar$). Consequently, the integral of the absolute difference provides only a loose upper bound: $\| \Delta W \| \le \int |W_1 - W_2| \, dz$.
Instead, the supremum definition in Eq.~\eqref{eq:op_norm_def} is mathematically equivalent to the \emph{Trace Distance} between the corresponding density operators $\hat{\rho}_1$ and $\hat{\rho}_2$. By the Helstrom-Holevo theorem:
\begin{equation}
    \| \Delta W \|^{\text{qm}} = \| \hat{\rho}_1 - \hat{\rho}_2 \|_{\text{tr}} \coloneqq \text{Tr}\left[ \sqrt{(\hat{\rho}_1 - \hat{\rho}_2)^\dagger (\hat{\rho}_1 - \hat{\rho}_2)} \right].
\end{equation}

\subsection{Fundamental Properties of the Base Norm}

To prove the proposition 2 regarding bounded deviation in the main text, we first establish the rigorous mathematical properties of the state space and the base norm. We begin by defining the structure of subsystems, then proceed to the properties of composition and evolution.

\subsubsection{ Structure of Subsystems and Marginalization}

Before analyzing the dynamics, we must formally establish the mathematical relationship between a composite system and its subsystems.  We demonstrate that the integral formulation of taking marginalization is the unique mathematical consequence of the probability definition and the structure of independent states established in the framework of Ref.~\cite{jiang2024framework}.

\renewcommand{\theproposition}{S\arabic{proposition}}
\setcounter{proposition}{0}

\begin{proposition}[Uniqueness of Phase Space Marginalization]\label{proposition_uniqueness_marginalization}
Let the measurement statistics be determined by the phase space integral pairing
\begin{equation}
    P(h|f) = \int_{\PS} h(z) f(z) \, d z \quad \text{(as defined in Eq.~17 of Ref.~\cite{jiang2024framework})},
\end{equation}
and let the joint state of an independent system be defined by the product structure
\begin{equation}
    W_{AB}(z_A, z_B) = W_A(z_A) \cdot W_B(z_B) \quad \text{(as defined in Eq.~79 of Ref.~\cite{jiang2024framework})},
\end{equation}
 where ``$\cdot$'' denotes the pointwise product of phase-space distributions.
Assuming statistical consistency (where local measurement statistics on subsystem $A$ are independent of the state of the discarded subsystem $B$, which is implicitly assumed in Ref.~\cite{jiang2024framework} and is a standard requirement for generalized probabilistic theories with no-signaling), the following hold:
\begin{enumerate}
    \item The effective measurement function $H(z_A, z_B)$ on the joint phase space corresponding to a local effect $h_A(z_A)$ is uniquely given by $H(z_A, z_B) = h_A(z_A) \cdot 1$.
    \item The marginal state $W_A$ is uniquely defined by the partial integration map $\mathcal{M}_B$:
    \begin{equation}
        W_A(z_A) = \mathcal{M}_B [W_{AB}](z_A) \coloneqq \int_{\PS_B} W_{AB}(z_A, z_B) \, d z_B.
    \end{equation}
\end{enumerate}
\end{proposition}

\begin{proof}
We rigorously derive the two claims sequentially, enforcing that the marginal state $W_A$ must reproduce the statistics of all local measurements on $A$.

\emph{Step 1: Characterization of local effects on the joint system.}
Fix a local effect $h_A : \PS_A \to \mathbb{R}$ on subsystem $A$. By Eq.~(17) of Ref.~\cite{jiang2024framework}, the probability of this outcome for a state $W_A$ is
\begin{equation}
    P(h_A|W_A) = \int_{\PS_A} h_A(z_A) W_A(z_A)\,dz_A.
\end{equation}

On the composite system $AB$, any outcome is described by some effect
$H : \PS_A \times \PS_B \to \mathbb{R}$.
We say that an effect $H$ on $AB$ \emph{implements the local effect $h_A$ on $A$ while ignoring $B$} if, for every product state $W_{AB} = W_A \cdot W_B$ (as in Eq.~(79) of Ref.~\cite{jiang2024framework}), it reproduces exactly the same statistics:
\begin{equation}\label{eq:stat_consistency_product}
    \int_{\PS_A} h_A(z_A) W_A(z_A)\,dz_A
    = \int_{\PS_{AB}} H(z_A,z_B)\, W_A(z_A) W_B(z_B)\,dz_A dz_B
\end{equation}
for all normalized $W_A \in \StateSpace_A$ and $W_B \in \StateSpace_B$.
In other words, describing the same physical experiment either as a measurement on the subsystem $A$ or as a measurement on the full system $AB$ must lead to identical probabilities for all uncorrelated preparations.

At this stage, Eq.~\eqref{eq:stat_consistency_product} should be read purely as a \emph{defining constraint} on those joint effects $H$ that qualify as implementations of the given local effect $h_A$ on the composite system. We are not requiring that every joint effect satisfy this relation.

It is straightforward to see that at least one such joint effect exists. Define
\begin{equation}\label{eq:H0_def}
    H_0(z_A,z_B) \coloneqq h_A(z_A)\,1_B(z_B),
\end{equation}
where $1_B(z_B)\equiv 1$ is the unit effect on subsystem $B$. Then, for any product state $W_{AB} = W_A \cdot W_B$,
\begin{align}
    \int_{\PS_{AB}} H_0(z_A,z_B)\, W_A(z_A) W_B(z_B)\,dz_A dz_B
    &= \int_{\PS_{AB}} h_A(z_A)\,1_B(z_B)\, W_A(z_A) W_B(z_B)\,dz_A dz_B \nonumber\\
    &= \left(\int_{\PS_A} h_A(z_A) W_A(z_A)\,dz_A\right)
       \left(\int_{\PS_B} W_B(z_B)\,dz_B\right) \nonumber\\
    &= \int_{\PS_A} h_A(z_A) W_A(z_A)\,dz_A,
\end{align}
since $W_B$ is normalized. Hence $H_0$ satisfies Eq.~\eqref{eq:stat_consistency_product} for all product states.

The remainder of Step~1 is devoted to showing that $H_0$ is in fact the \emph{unique} joint effect with this property: if $H$ obeys Eq.~\eqref{eq:stat_consistency_product} for all normalized $W_A$ and $W_B$, then $H(z_A,z_B) = H_0(z_A,z_B)$ for almost all $(z_A,z_B)\in\PS_A\times\PS_B$.

Let $H$ be any joint effect on $\PS_{AB}$ that satisfies Eq.~\eqref{eq:stat_consistency_product} for all normalized $W_A \in \StateSpace_A$ and $W_B \in \StateSpace_B$. Fix an arbitrary normalized $W_B$ and apply Fubini's theorem to the right-hand side of Eq.~\eqref{eq:stat_consistency_product}:
\begin{align}
    \int_{\PS_{AB}} H(z_A,z_B)\, W_A(z_A) W_B(z_B)\,dz_A dz_B
    &= \int_{\PS_A} W_A(z_A)\left[\int_{\PS_B} H(z_A,z_B) W_B(z_B)\,dz_B\right]dz_A.
\end{align}
Thus Eq.~\eqref{eq:stat_consistency_product} can be rewritten as
\begin{equation}\label{eq:WA_identity_again}
    \int_{\PS_A} h_A(z_A) W_A(z_A)\,dz_A
    = \int_{\PS_A} W_A(z_A)
      \left[\int_{\PS_B} H(z_A,z_B) W_B(z_B)\,dz_B\right]dz_A
\end{equation}
for all normalized $W_A \in \StateSpace_A$, with the chosen $W_B$ held fixed.

Define, for this fixed $W_B$, the function
\begin{equation}
    F_{W_B}(z_A) \coloneqq
    h_A(z_A) - \int_{\PS_B} H(z_A,z_B) W_B(z_B)\,dz_B.
\end{equation}
Then \eqref{eq:WA_identity_again} is equivalent to
\begin{equation}\label{eq:F_WB_zero_pairing}
    \int_{\PS_A} F_{W_B}(z_A)\, W_A(z_A)\,dz_A = 0
    \quad \text{for all normalized } W_A \in \StateSpace_A.
\end{equation}

Here we invoke a standard GPT assumption, compatible with the generalized phase-space framework of Ref.~\cite{jiang2024framework}: \emph{states separate effects}. Concretely, if a function $F$ satisfies
\[
    \int_{\PS_A} F(z_A)\, W_A(z_A)\,dz_A = 0
\]
for all normalized $W_A \in \StateSpace_A$, then the corresponding effect is operationally indistinguishable from the zero effect, and we identify $F(z_A)=0$ almost everywhere on $\PS_A$. Applying this separating property to \eqref{eq:F_WB_zero_pairing}, we obtain
\begin{equation}\label{eq:H_constraint_WB}
    h_A(z_A)
    =
    \int_{\PS_B} H(z_A,z_B) W_B(z_B)\,dz_B
\end{equation}
for almost all $z_A \in \PS_A$, for the fixed but arbitrary normalized $W_B \in \StateSpace_B$.

We now use the arbitrariness of $W_B$ to constrain the $z_B$-dependence of $H$. For each fixed $z_A$, define a linear functional (an effect on subsystem $B$)
\begin{equation}
    e_{z_A}(W_B) \coloneqq \int_{\PS_B} H(z_A,z_B)\,W_B(z_B)\,dz_B,
    \qquad W_B \in \StateSpace_B.
\end{equation}
Equation~\eqref{eq:H_constraint_WB} states that, for every normalized $W_B$,
\[
    e_{z_A}(W_B) = h_A(z_A),
\]
i.e.\ $e_{z_A}$ is a constant functional on the entire state space of $B$, with value $h_A(z_A)$.

By the usual GPT structure, there is a distinguished \emph{unit effect} $1_B$ on subsystem $B$, characterized by
\[
    1_B(W_B) = 1 \quad \text{for all normalized } W_B \in \StateSpace_B,
\]
and this unit effect is unique. Therefore, for each fixed $z_A$, the only effect on $B$ that takes the constant value $h_A(z_A)$ on all normalized states is the scalar multiple $h_A(z_A)\,1_B$. In other words, as effects on $B$ we have
\[
    e_{z_A} = h_A(z_A)\,1_B.
\]

Recalling that in the generalized phase-space representation every effect on $B$ is given by a real function $h_B(z_B)$ via
\[
    e(W_B) = \int_{\PS_B} h_B(z_B)\,W_B(z_B)\,dz_B,
\]
we conclude that, for almost all $(z_A,z_B)\in\PS_A\times\PS_B$,
\begin{equation}
    H(z_A,z_B) = h_A(z_A)\,1_B(z_B).
\end{equation}
Equivalently,
\begin{equation}
    H(z_A,z_B) = h_A(z_A)\cdot 1_B(z_B) = H_0(z_A,z_B),
\end{equation}
where $1_B(z_B)\equiv 1$ is the unit effect on subsystem $B$ and $H_0$ is the joint effect defined in Eq.~\eqref{eq:H0_def}.

This shows that any joint effect $H$ that satisfies the statistical consistency condition \eqref{eq:stat_consistency_product} for all product states $W_A \cdot W_B$ must coincide almost everywhere with $H_0$. In particular, $H_0$ is the unique implementation of the local effect $h_A$ on the composite system $AB$, which completes Step~1.

\emph{Step 2: Derivation of the marginalization map.}
We now consider a general (possibly correlated) joint state
$W_{AB} \in \StateSpace_{AB}$.
By definition, a state $W_A$ on subsystem $A$ is a \emph{marginal} of $W_{AB}$ if it reproduces the statistics of all local measurements on $A$. That is, for every local effect $h_A : \PS_A \to \mathbb{R}$, the probability computed using $W_A$ must coincide with the probability obtained from the joint description using the corresponding effect $H = h_A \cdot 1_B$ on $AB$:
\begin{equation}\label{eq:marginal_def}
    \int_{\PS_A} h_A(z_A) W_A(z_A)\,dz_A
    = \int_{\PS_{AB}} h_A(z_A)\,1_B(z_B)\, W_{AB}(z_A,z_B)\,dz_A dz_B
\end{equation}
for all effects $h_A$ on $A$. This is the precise statement that the reduced description $W_A$ yields the same measurement statistics as the full description $W_{AB}$ for all measurements that act only on $A$ and ignore $B$.

We first show that the usual partial-integration formula provides a marginal in this sense. Define a candidate reduced state
\begin{equation}\label{eq:marginal_candidate}
    \widetilde{W}_A(z_A)
    \coloneqq
    \int_{\PS_B} W_{AB}(z_A,z_B)\,dz_B.
\end{equation}
Then, for any local effect $h_A$ on $A$, we have
\begin{align}
    \int_{\PS_A} h_A(z_A)\,\widetilde{W}_A(z_A)\,dz_A
    &= \int_{\PS_A} h_A(z_A)
       \left[\int_{\PS_B} W_{AB}(z_A,z_B)\,dz_B\right]dz_A \nonumber\\
    &= \int_{\PS_{AB}} h_A(z_A)\, W_{AB}(z_A,z_B)\,dz_A dz_B,
\end{align}
where we used Fubini's theorem to interchange the order of integration. Since $1_B(z_B)\equiv 1$, this coincides with the right-hand side of \eqref{eq:marginal_def} for $H = h_A \cdot 1_B$. Thus $\widetilde{W}_A$ as defined in \eqref{eq:marginal_candidate} indeed reproduces the statistics of all local measurements on $A$ and therefore is a marginal of $W_{AB}$.

It remains to prove that this marginal is unique. Suppose $W_A$ and $W_A'$ are two states on $\PS_A$ that both satisfy the marginality condition \eqref{eq:marginal_def} for the same joint state $W_{AB}$. Then, for every local effect $h_A$, we have
\begin{equation}
    \int_{\PS_A} h_A(z_A)\, W_A(z_A)\,dz_A
    = \int_{\PS_{AB}} h_A(z_A)\, W_{AB}(z_A,z_B)\,dz_A dz_B
    = \int_{\PS_A} h_A(z_A)\, W_A'(z_A)\,dz_A.
\end{equation}
Equivalently,
\begin{equation}\label{eq:WA_diff_pairing}
    \int_{\PS_A} h_A(z_A)\,\bigl(W_A(z_A) - W_A'(z_A)\bigr)\,dz_A = 0
    \quad \text{for all local effects } h_A.
\end{equation}

By the axioms of the generalized phase-space framework (Sec.~II.C of Ref.~\cite{jiang2024framework}), effects separate states: if a difference $f \in \VecSpace_A$ satisfies
\[
    \int_{\PS_A} h_A(z_A)\, f(z_A)\,dz_A = 0
\]
for all effects $h_A$ on $A$, then $f(z_A) = 0$ almost everywhere on $\PS_A$. (This is exactly the statement that the ordered vector space of states is separated by its dual cone of effects.) Applying this separation property to the difference
\[
    f(z_A) \coloneqq W_A(z_A) - W_A'(z_A),
\]
and using \eqref{eq:WA_diff_pairing}, we conclude that
\begin{equation}
    W_A(z_A) = W_A'(z_A)
\end{equation}
for almost all $z_A \in \PS_A$.

Therefore, the map that assigns to each joint state $W_{AB}$ the function
\begin{equation}
    W_A(z_A) = \mathcal{M}_B[W_{AB}](z_A)
    \coloneqq \int_{\PS_B} W_{AB}(z_A,z_B)\,dz_B
\end{equation}
is the \emph{unique} marginalization map from $\StateSpace_{AB}$ to $\StateSpace_A$ that reproduces the statistics of all local measurements on $A$.

\end{proof}

\setcounter{theorem}{0}
\renewcommand{\thetheorem}{S\arabic{theorem}}

\begin{lemma}[Contractivity of marginalization]
The distinguishability of marginal states cannot exceed the distinguishability of their joint states:
\begin{equation}
    \bigl\| \mathcal{M}_B[W_{AB}] \bigr\|
    \;\le\;
    \| W_{AB} \|.
\end{equation}
\label{lem:marginal_contractivity}
\end{lemma}

\begin{proof}
Let $W_A \coloneqq \mathcal{M}_B[W_{AB}]$ denote the marginal state on $A$, defined by
\begin{equation}
    W_A(z_A) = \int_{\PS_B} W_{AB}(z_A,z_B)\,dz_B.
\end{equation}
By definition of the base norm on subsystem $A$, we have
\begin{equation}\label{eq:base_norm_A}
    \|W_A\|
    =
    \sup_{\{h_{A,i}\}}
    \sum_i \left| \int_{\PS_A} h_{A,i}(z_A)\,W_A(z_A)\,dz_A \right|,
\end{equation}
where the supremum is taken over all finite (or countable) collections of effects $\{h_{A,i}\}$ on $A$ that form a complete measurement, i.e.\ $\sum_i h_{A,i} = 1_A$.

For any such measurement $\{h_{A,i}\}$ on $A$, consider the corresponding \emph{local} measurement on $AB$ defined by
\begin{equation}
    H_i(z_A,z_B) \coloneqq h_{A,i}(z_A)\,1_B(z_B),
\end{equation}
where $1_B(z_B)\equiv 1$ is the unit effect on subsystem $B$. By Step~1 above, each $H_i$ is a valid effect on $AB$, and the collection $\{H_i\}$ is complete on $AB$ since
\[
    \sum_i H_i(z_A,z_B)
    = \sum_i h_{A,i}(z_A)\,1_B(z_B)
    = 1_A(z_A)\,1_B(z_B)
    = 1_{AB}(z_A,z_B).
\]

Using the definition of the marginal and Fubini's theorem, we compute, for each $i$,
\begin{align}
    \int_{\PS_A} h_{A,i}(z_A)\,W_A(z_A)\,dz_A
    &= \int_{\PS_A} h_{A,i}(z_A)
       \left[\int_{\PS_B} W_{AB}(z_A,z_B)\,dz_B\right]dz_A \nonumber\\
    &= \int_{\PS_{AB}} h_{A,i}(z_A)\,1_B(z_B)\,W_{AB}(z_A,z_B)\,dz_A dz_B \nonumber\\
    &= \int_{\PS_{AB}} H_i(z_A,z_B)\,W_{AB}(z_A,z_B)\,dz_A dz_B.
\end{align}
Substituting this into \eqref{eq:base_norm_A}, we obtain
\begin{equation}
    \|W_A\|
    =
    \sup_{\{h_{A,i}\}}
    \sum_i \left| \int_{\PS_{AB}} H_i(z_A,z_B)\,W_{AB}(z_A,z_B)\,dz_A dz_B \right|.
\end{equation}

Now observe that every family $\{H_i\}$ constructed in this way is a valid complete measurement on $AB$, but such families form only a \emph{subset} of all possible measurements on $AB$ (they are precisely those that act trivially on $B$). By definition of the base norm on $AB$,
\begin{equation}
    \|W_{AB}\|
    =
    \sup_{\{G_j\}}
    \sum_j \left| \int_{\PS_{AB}} G_j(z_A,z_B)\,W_{AB}(z_A,z_B)\,dz_A dz_B \right|,
\end{equation}
where the supremum is taken over \emph{all} complete measurements $\{G_j\}$ on $AB$.

Since the supremum over a subset of measurements cannot exceed the supremum over the entire set, we have
\begin{equation}
    \|W_A\|
    =
    \sup_{\{H_i \text{ of the form } h_{A,i}\cdot 1_B\}}
    \sum_i \left| \int_{\PS_{AB}} H_i W_{AB} \right|
    \;\le\;
    \sup_{\{G_j\}} \sum_j \left| \int_{\PS_{AB}} G_j W_{AB} \right|
    = \|W_{AB}\|.
\end{equation}
This is exactly the desired inequality
\(
    \|\mathcal{M}_B[W_{AB}]\| \le \|W_{AB}\|.
\)
\end{proof}

\subsubsection{ Composition of base norm (product states)}

We first address the structure of composite systems. Consider a joint system composed of the Harvester ($A$) and the Source ($B$).
\begin{itemize}
    \item The joint phase space is the Cartesian product $\PS_{AB} = \PS_A \times \PS_B$.
    \item Adhering to the definition of uncorrelated subsystems provided in Eq.~(79) of Ref.~\cite{jiang2024framework}, the joint state of an independent system is given by the product of the marginals:
    \begin{equation}
        W_{AB}(z_A,z_B) = W_A(z_A)\cdot W_B(z_B),
    \end{equation}
    where ``$\cdot$'' denotes the pointwise product of phase-space distributions.
\end{itemize}

\begin{lemma}[Composition of base norm for uncorrelated states]\label{lem:product_norm}
Let $W_A \in \StateSpace_A$ be a fixed, normalized physical state of system $A$, and let $\Delta W_B \in \VecSpace_B$ be a difference vector for system $B$. Under the structural assumption that independent composite states are represented by the product of their marginals (as in Eq.~(79) of Ref.~\cite{jiang2024framework}), the base norm satisfies
\begin{equation}
    \bigl\| W_A \cdot \Delta W_B \bigr\| = \bigl\| \Delta W_B \bigr\|.
\end{equation}
\end{lemma}

\emph{Physical interpretation.} Adding an uncorrelated auxiliary system $A$ in a known state $W_A$ does not change the operational distinguishability of the states on system $B$.

\begin{proof}
Set
\[
    v_{AB}(z_A,z_B) \coloneqq W_A(z_A)\,\Delta W_B(z_B).
\]
By construction of the generalized phase-space vector spaces (Sec.~II.C of Ref.~\cite{jiang2024framework}), $v_{AB}$ is an admissible vector in $\VecSpace_{AB}$, so its base norm is well-defined. We prove the claimed equality by establishing both the lower bound ($\ge$) and the upper bound ($\le$).

\medskip\noindent
\emph{Step 1: Lower bound $\bigl\|v_{AB}\bigr\| \ge \bigl\|\Delta W_B\bigr\|$.}
The base norm on the joint system $AB$ is defined as the supremum over \emph{all} complete measurements $\{H_i\}$ on $AB$:
\begin{equation}
    \|v_{AB}\|
    = \sup_{\{H_i\}\subset\mathcal{E}_{AB}}
      \sum_i \left| \int_{\PS_{AB}} H_i(z_A,z_B)\,v_{AB}(z_A,z_B)\,dz_A dz_B \right|,
\end{equation}
where $\mathcal{E}_{AB}$ denotes all effects on the compsite system $AB$.
In particular, consider the subset of joint measurements that act non-trivially only on $B$ and trivially on $A$. Let $\{h_{B,i}(z_B)\}$ be an arbitrary complete measurement on system $B$. By the same reasoning as in Proposition~\ref{proposition_uniqueness_marginalization} (applied with $A$ and $B$ interchanged), the joint effect on $\PS_{AB}$ that implements the local effect $h_{B,i}$ on $B$ while ignoring $A$ is
\begin{equation}
    H_i(z_A,z_B) \coloneqq 1_A(z_A)\,h_{B,i}(z_B),
\end{equation}
where $1_A(z_A)\equiv 1$ is the unit effect on subsystem $A$. The collection $\{H_i\}$ is a valid complete measurement on $AB$.

For this particular joint measurement, the corresponding contribution to the norm of $v_{AB}$ is
\begin{align}
    S(\{H_i\})
    &\coloneqq
    \sum_i \left| \int_{\PS_{AB}} H_i(z_A,z_B)\,v_{AB}(z_A,z_B)\,dz_A dz_B \right| \nonumber\\
    &= \sum_i \left|
       \int_{\PS_A}\int_{\PS_B}
           1_A(z_A)\,h_{B,i}(z_B)\,W_A(z_A)\,\Delta W_B(z_B)\,dz_B dz_A
       \right| \nonumber\\
    &= \sum_i \left|
       \left( \int_{\PS_A} W_A(z_A)\,dz_A \right)
       \left( \int_{\PS_B} h_{B,i}(z_B)\,\Delta W_B(z_B)\,dz_B \right)
       \right|.
\end{align}
Using the normalization of $W_A$, $\int_{\PS_A} W_A(z_A)\,dz_A = 1$, this simplifies to
\begin{equation}
    S(\{H_i\}) = \sum_i \left| \int_{\PS_B} h_{B,i}(z_B)\,\Delta W_B(z_B)\,dz_B \right|.
\end{equation}
Taking the supremum over all measurements $\{h_{B,i}\}$ on $B$, we obtain
\begin{equation}
    \sup_{\{h_{B,i}\}\subset\mathcal{E}_B} S(\{H_i\})
    = \bigl\|\Delta W_B\bigr\|.
\end{equation}
Since the base norm on $AB$ is the supremum over a \emph{larger} set of measurements (it includes all such $\{H_i\}$ and possibly more general, correlated measurements), we conclude that
\begin{equation}
    \bigl\|v_{AB}\bigr\|
    \ge \sup_{\{h_{B,i}\}\subset\mathcal{E}_B} S(\{H_i\})
    = \bigl\|\Delta W_B\bigr\|.
\end{equation}

\medskip\noindent
\emph{Step 2: Upper bound $\bigl\|v_{AB}\bigr\| \le \bigl\|\Delta W_B\bigr\|$.}
We now show that no joint measurement on $AB$ can extract more statistical distinguishability from $v_{AB}$ than is already present in $\Delta W_B$ alone.

Let $\{H_k(z_A,z_B)\}$ be an arbitrary complete measurement on $AB$. For this fixed measurement, define
\begin{align}\label{eq:S_Hk_def}
    S(\{H_k\})
    &\coloneqq
    \sum_k \left| \int_{\PS_{AB}} H_k(z_A,z_B)\,v_{AB}(z_A,z_B)\,dz_A dz_B \right| \\
   & =
    \sum_k \left| \int_{\PS_{AB}} H_k(z_A,z_B)\,W_A(z_A)\,\Delta W_B(z_B)\,dz_A dz_B \right|.
\end{align}
By definition of the base norm on $AB$,
\begin{equation}
    \bigl\|v_{AB}\bigr\|
    = \sup_{\{H_k\}\subset\mathcal{E}_{AB}} S(\{H_k\}).
\end{equation}

Using Fubini's theorem, we first integrate over $z_A$ in \eqref{eq:S_Hk_def} and define, for each $k$, the \emph{induced effect} on $B$:
\begin{equation}\label{eq:induced_effect_def}
    \tilde{h}_k(z_B)
    \coloneqq
    \int_{\PS_A} H_k(z_A,z_B)\,W_A(z_A)\,dz_A.
\end{equation}
Then we can rewrite
\begin{equation}
    S(\{H_k\})
    = \sum_k \left| \int_{\PS_B} \tilde{h}_k(z_B)\,\Delta W_B(z_B)\,dz_B \right|.
\end{equation}

We now verify that the collection $\{\tilde{h}_k\}$ is a valid complete measurement on subsystem $B$.

\smallskip\noindent
\emph{(i) Each $\tilde{h}_k$ is a valid effect on $B$.}
Let $\rho_B(z_B)$ be any valid state of system $B$. Consider the product state
\[
    W_{AB}^{(\rho_B)}(z_A,z_B) \coloneqq W_A(z_A)\,\rho_B(z_B).
\]
By the independence postulate (Eq.~(79) in Ref.~\cite{jiang2024framework}), this is a valid state on $AB$. The probability of outcome $k$ when measuring $\{H_k\}$ on $W_{AB}^{(\rho_B)}$ is
\begin{align}
    P(k)
    &= \int_{\PS_{AB}} H_k(z_A,z_B)\,W_A(z_A)\,\rho_B(z_B)\,dz_A dz_B \nonumber\\
    &= \int_{\PS_B} \left[ \int_{\PS_A} H_k(z_A,z_B)\,W_A(z_A)\,dz_A \right] \rho_B(z_B)\,dz_B \nonumber\\
    &= \int_{\PS_B} \tilde{h}_k(z_B)\,\rho_B(z_B)\,dz_B.
\end{align}
Since $H_k$ is a valid effect on $AB$ and $W_{AB}^{(\rho_B)}$ is a valid state, we must have $0 \le P(k) \le 1$ for all $\rho_B$. Hence $\tilde{h}_k$ maps every valid state of $B$ to a probability in $[0,1]$, and is therefore a valid effect on $B$.

\smallskip\noindent
\emph{(ii) The set $\{\tilde{h}_k\}$ is complete on $B$.}
Using the completeness of $\{H_k\}$ on $AB$, $\sum_k H_k(z_A,z_B) = 1_{AB}(z_A,z_B) \equiv 1$, we obtain
\begin{align}
    \sum_k \tilde{h}_k(z_B)
    &= \sum_k \int_{\PS_A} H_k(z_A,z_B)\,W_A(z_A)\,dz_A \nonumber\\
    &= \int_{\PS_A} \left[ \sum_k H_k(z_A,z_B) \right] W_A(z_A)\,dz_A \nonumber\\
    &= \int_{\PS_A} 1_{AB}(z_A,z_B)\,W_A(z_A)\,dz_A \nonumber\\
    &= 1_B(z_B) \int_{\PS_A} W_A(z_A)\,dz_A.
\end{align}
Here $1_B(z_B)\equiv 1$ is the unit effect on $B$. Since $W_A$ is normalized, $\int_{\PS_A} W_A(z_A)\,dz_A = 1$, so
\begin{equation}
    \sum_k \tilde{h}_k(z_B) = 1_B(z_B).
\end{equation}
Thus $\{\tilde{h}_k\}$ is a complete measurement on $B$.

\smallskip

Having established that $\{\tilde{h}_k\}$ is a valid complete measurement on $B$, we can now bound $S(\{H_k\})$ using the definition of the base norm on $B$:
\begin{equation}
    S(\{H_k\})
    = \sum_k \left| \int_{\PS_B} \tilde{h}_k(z_B)\,\Delta W_B(z_B)\,dz_B \right|
    \le \sup_{\{g_i\}\subset\mathcal{E}_B}
        \sum_i \left| \int_{\PS_B} g_i(z_B)\,\Delta W_B(z_B)\,dz_B \right|
    = \bigl\|\Delta W_B\bigr\|.
\end{equation}
Since $\{H_k\}$ was an arbitrary complete measurement on $AB$, this inequality holds for every choice of $\{H_k\}$. Hence, taking the supremum over all joint measurements,
\begin{equation}
    \bigl\|v_{AB}\bigr\|
    = \sup_{\{H_k\}\subset\mathcal{E}_{AB}} S(\{H_k\})
    \le \bigl\|\Delta W_B\bigr\|.
\end{equation}

\medskip\noindent
\emph{Conclusion.}
Combining the lower bound from Step~1 and the upper bound from Step~2, we obtain
\begin{equation}
    \bigl\| W_A \cdot \Delta W_B \bigr\|
    = \bigl\| \Delta W_B \bigr\|.
\end{equation}
This completes the proof.
\end{proof}

\subsubsection{Isometry under reversible evolution}

We consider the time evolution of a single system from $t=0$ to $t=\tau$. Let
\[
    \Phi_\tau : \VecSpace \to \VecSpace
\]
be the linear map representing this evolution, so that $W(\tau) = \Phi_\tau[W(0)]$.

In the framework of Ref.~\cite{jiang2024framework} (Sec.~VI), the dynamics are generated by a generalized Hamiltonian. According to Theorem~5 of Ref.~\cite{jiang2024framework}, the time evolution of the state $W(z,t)$ is governed by the generalized Liouville equation
\begin{equation}
    \frac{\partial W}{\partial t}
    = \mathcal{L}_H W
    \coloneqq \mathcal{N} \int K(k)\,
        \sin\!\left( \frac{k}{2}\Lambda \right)(W,H)\,dk,
\end{equation}
where $H(z)$ is the Hamiltonian function defined by the pure stationary states of the system, $\Lambda$ is the symplectic operator (Poisson bracket), $K(k)$ is the theory-specific nonlocal dynamics kernel, and $\mathcal{N}$ is a normalization constant. The notation $\sin(\tfrac{k}{2}\Lambda)(W,H)$ denotes the action of the corresponding pseudodifferential operator series on the pair $(W,H)$.

Since the Hamiltonian $H$ is time-independent (due to time translation symmetry, Postulate~1 in Ref.~\cite{jiang2024framework}), the generator $\mathcal{L}_H$ is a linear, time-independent operator on $\VecSpace$. Consequently, the finite-time evolution map is formally given by the exponential
\[
    \Phi_\tau = \exp(\mathcal{L}_H \tau),
\]
and the family $\{\Phi_\tau\}_{\tau\in\mathbb{R}}$ forms a one-parameter group:
\[
    \Phi_{\tau+\sigma} = \Phi_\tau \circ \Phi_\sigma,
    \qquad
    \Phi_0 = \id,
    \qquad
    \Phi_\tau^{-1} = \Phi_{-\tau}.
\]
For the lemma below, we only use that $\Phi_\tau$ is a linear bijection on $\VecSpace$ which preserves probabilities and the bilinear form (inner product) specified in Postulate~5 of Ref.~\cite{jiang2024framework}.

\begin{lemma}[Isometry] \label{lem:isometry}
Under any reversible, probability-preserving linear evolution $\Phi_\tau$, the base norm is invariant:
\begin{equation}
    \bigl\| \Phi_\tau v \bigr\| = \| v \|
    \qquad \text{for all } v \in \VecSpace.
\end{equation}
\end{lemma}

\begin{proof}
We split the argument into two steps.

\medskip\noindent
\emph{Step 1: The adjoint equals the inverse.}
Postulate~5 of Ref.~\cite{jiang2024framework} asserts that the bilinear pairing
\[
    \langle f,g\rangle \coloneqq \int_{\PS} f(z)\,g(z)\,dz
\]
is invariant under the dynamics:
\begin{equation}\label{eq:inner_product_invariance}
    \int_{\PS} (\Phi_\tau f)(z)\,(\Phi_\tau g)(z)\,dz
    = \int_{\PS} f(z)\,g(z)\,dz
\end{equation}
for all (sufficiently regular) real functions $f,g$ on $\PS$. In operator language, this means that $\Phi_\tau$ is an isometry with respect to this inner product.

By definition of the adjoint operator $\Phi_\tau^*$ with respect to this pairing, we have
\begin{equation}\label{eq:adjoint_def}
    \int_{\PS} (\Phi_\tau u)(z)\,v(z)\,dz
    = \int_{\PS} u(z)\,(\Phi_\tau^* v)(z)\,dz
    \qquad\text{for all } u,v \in \VecSpace.
\end{equation}
Apply \eqref{eq:adjoint_def} to $u=f$ and $v=\Phi_\tau g$:
\begin{equation}
    \int_{\PS} (\Phi_\tau f)(z)\,(\Phi_\tau g)(z)\,dz
    = \int_{\PS} f(z)\,(\Phi_\tau^* \Phi_\tau g)(z)\,dz.
\end{equation}
Comparing this with \eqref{eq:inner_product_invariance}, we obtain
\begin{equation}
    \int_{\PS} f(z)\,(\Phi_\tau^* \Phi_\tau g)(z)\,dz
    = \int_{\PS} f(z)\,g(z)\,dz
\end{equation}
for all $f,g$. Equivalently,
\begin{equation}
    \int_{\PS} f(z)\,\bigl[(\Phi_\tau^* \Phi_\tau g)(z) - g(z)\bigr]\,dz = 0
    \qquad\text{for all } f,g.
\end{equation}
This implies
\begin{equation}\label{eq:Phi_star_Phi_id}
    \Phi_\tau^* \Phi_\tau = \id
\end{equation}
as operators on $\VecSpace$.

On the other hand, by reversibility (Postulate~1 of Ref.~\cite{jiang2024framework}) the map $\Phi_\tau$ is invertible with inverse $\Phi_\tau^{-1} = \Phi_{-\tau}$. Comparing \eqref{eq:Phi_star_Phi_id} with $\Phi_\tau^{-1} \Phi_\tau = \id$ and using uniqueness of inverses, we conclude that
\begin{equation}\label{eq:adjoint_equals_inverse}
    \Phi_\tau^* = \Phi_\tau^{-1} = \Phi_{-\tau}.
\end{equation}

\medskip\noindent
\emph{Step 2: Invariance of the base norm.}
Let $v \in \VecSpace$ be arbitrary, and set $v(\tau) \coloneqq \Phi_\tau v$. Denote by $\mathcal{E}$ the set of all valid effects on the system, and by $\mathfrak{M}$ the set of all complete measurements, i.e.\ all finite (or countable) families $\{h_i\} \subset \mathcal{E}$ such that $\sum_i h_i = 1$.

By definition of the base norm,
\begin{equation}\label{eq:base_norm_evolved}
    \| v(\tau) \|
    = \sup_{\{h_i\}\in\mathfrak{M}}
      \sum_i \left| \int_{\PS} h_i(z)\,(\Phi_\tau v)(z)\,dz \right|.
\end{equation}
Using the adjoint relation \eqref{eq:adjoint_def} with $u=v$ and $v=h_i$, and then substituting \eqref{eq:adjoint_equals_inverse}, we obtain
\begin{equation}
    \int_{\PS} h_i(z)\,(\Phi_\tau v)(z)\,dz
    = \int_{\PS} (\Phi_\tau^* h_i)(z)\,v(z)\,dz
    = \int_{\PS} (\Phi_{-\tau} h_i)(z)\,v(z)\,dz.
\end{equation}
Therefore
\begin{equation}
    \| v(\tau) \|
    = \sup_{\{h_i\}\in\mathfrak{M}}
      \sum_i \left| \int_{\PS} (\Phi_{-\tau} h_i)(z)\,v(z)\,dz \right|.
\end{equation}

Define, for each measurement $\{h_i\}\in\mathfrak{M}$, a new family of functions $\{g_i\}$ by
\begin{equation}\label{eq:gi_def}
    g_i \coloneqq \Phi_{-\tau} h_i.
\end{equation}
We now verify that this construction defines a bijection on $\mathfrak{M}$.

\smallskip\noindent
\emph{(i) $\Phi_{-\tau}$ maps effects to effects.}
Let $h \in \mathcal{E}$ be any valid effect and define $g = \Phi_{-\tau} h$. For any valid state $W$, the probability associated with $g$ is
\begin{align}
    \int_{\PS} g(z)\,W(z)\,dz
    &= \int_{\PS} (\Phi_{-\tau} h)(z)\,W(z)\,dz \nonumber\\
    &= \int_{\PS} h(z)\,(\Phi_\tau W)(z)\,dz,
\end{align}
where we used the adjoint relation with $u=W$ and $v=h$. Since $\Phi_\tau$ maps valid states to valid states and $h$ is a valid effect, the right-hand side lies in $[0,1]$ for all $W$. Thus $g$ is again a valid effect, i.e.\ $g \in \mathcal{E}$.

\smallskip\noindent
\emph{(ii) Completeness is preserved.}
If $\{h_i\}\in\mathfrak{M}$ is a complete measurement on $\PS$, then $\sum_i h_i = 1$. Applying $\Phi_{-\tau}$ and using linearity,
\begin{equation}
    \sum_i g_i
    = \sum_i \Phi_{-\tau} h_i
    = \Phi_{-\tau} \left( \sum_i h_i \right)
    = \Phi_{-\tau}(1).
\end{equation}
Because $\Phi_\tau$ preserves total probability, its dual map $\Phi_{-\tau}$ must preserve the unit effect: for any state $W$,
\[
    \int_{\PS} \Phi_{-\tau}(1)(z)\,W(z)\,dz
    = \int_{\PS} 1(z)\,(\Phi_\tau W)(z)\,dz
    = 1.
\]
By uniqueness of the unit effect, this implies $\Phi_{-\tau}(1) = 1$, and therefore $\sum_i g_i = 1$. Hence $\{g_i\}\in\mathfrak{M}$.

\smallskip\noindent
\emph{(iii) Bijection on $\mathfrak{M}$.}
Since $\Phi_{-\tau}$ is invertible with inverse $\Phi_\tau$, the map $\{h_i\}\mapsto\{g_i=\Phi_{-\tau}h_i\}$ is a bijection from $\mathfrak{M}$ onto itself. Therefore, the supremum over all $\{h_i\}\in\mathfrak{M}$ is equal to the supremum over all $\{g_i\}\in\mathfrak{M}$.

Substituting $g_i = \Phi_{-\tau}h_i$ into the expression for the norm, and using this bijection, we obtain
\begin{equation}
    \| v(\tau) \|
    = \sup_{\{h_i\}\in\mathfrak{M}}
      \sum_i \left| \int_{\PS} (\Phi_{-\tau} h_i)(z)\,v(z)\,dz \right|
    = \sup_{\{g_i\}\in\mathfrak{M}}
      \sum_i \left| \int_{\PS} g_i(z)\,v(z)\,dz \right|.
\end{equation}
The right-hand side is precisely the definition of the base norm of $v$ at time $t=0$:
\begin{equation}
    \| v(\tau) \| = \| v \|.
\end{equation}

This shows that the base norm is invariant under any reversible, probability-preserving linear evolution $\Phi_\tau$, i.e.\ the operational distinguishability of states is conserved under generalized Hamiltonian dynamics.
\end{proof}

\subsection{ Proof the distinguishability bound for the harvester state (Proposition 2 in main text)}

We now assemble the three structural lemmas (contractivity under marginalization---Lemma~\ref{lem:marginal_contractivity}), product composition of the base norm---Lemma~\ref{lem:product_norm} and isometry under reversible evolution---Lemma~\ref{lem:isometry} ) to prove the main bound of Proposition 2 in the main text.

\renewcommand{\theproposition}{\arabic{proposition}}
\setcounter{proposition}{1}

\begin{proposition}[ Distinguishability bound for the harvester state]
Consider a composite system initialized in an uncorrelated product state. Let the harvester ($A$) be in a fixed state $W_A^{(0)}$, and the source ($B$) be in one of two possible states, $W_{B,1}^{(0)}$ or $W_{B,2}^{(0)}$. Let $\Delta W_B(0) \coloneqq W_{B,1}^{(0)} - W_{B,2}^{(0)}$ denote the initial source perturbation.

Under any reversible, probability-preserving linear evolution
$\Phi_\tau^{AB}$ allowed by the generalized Hamiltonian
framework of Ref.~\cite{jiang2024framework}, the deviation in the
harvester's state at time $\tau$, denoted $\Delta W_A(\tau)$,
satisfies the bound
\begin{equation}
    \bigl\| \Delta W_A(\tau) \bigr\|
    \;\le\;
    \bigl\| \Delta W_B(0) \bigr\|.
\end{equation}
Using the identification $D(W_1,W_2) = \|W_1 - W_2\|$, this is precisely the inequality stated in Eq.~(5) of the main text for the case of two source states $W_{B,1},W_{B,2}$ and their corresponding harvester outputs $W_{A,1}(\tau),W_{A,2}(\tau)$.
\end{proposition}

\begin{proof}
\textit{Step 1: Initialization on the composite system.}
The two possible initial joint states of $AB$ are
\[
    W_{AB,1}^{(0)}(z_A,z_B)
    = W_A^{(0)}(z_A)\,W_{B,1}^{(0)}(z_B),
    \qquad
    W_{AB,2}^{(0)}(z_A,z_B)
    = W_A^{(0)}(z_A)\,W_{B,2}^{(0)}(z_B),
\]
so their difference is
\begin{equation}
\label{eq:Delta_WAB_0}
    \Delta W_{AB}(0)
    \coloneqq W_{AB,1}^{(0)} - W_{AB,2}^{(0)}
    = W_A^{(0)}(z_A)\,\Delta W_B(0)(z_B).
\end{equation}
Since $W_A^{(0)}$ is a valid normalized state of $A$ ($\int_{\PS_A}W_A^{(0)}=1$), we can apply the lemma on the composition of the base norm for uncorrelated states (Lemma~\ref{lem:product_norm}) to obtain
\begin{equation}
    \bigl\| \Delta W_{AB}(0) \bigr\|
    = \bigl\| W_A^{(0)} \cdot \Delta W_B(0) \bigr\|
    = \bigl\| \Delta W_B(0) \bigr\|.
\end{equation}

\textit{Step 2: Reversible evolution preserves total distinguishability.}
Let $\Phi_\tau^{AB}$ denote the generalized Hamiltonian evolution on the composite system $AB$ from time $0$ to $\tau$. The evolved difference vector is
\begin{equation}
    \Delta W_{AB}(\tau)
    \coloneqq \Phi_\tau^{AB}\bigl[ \Delta W_{AB}(0) \bigr].
\end{equation}
By the isometry lemma (Lemma~\ref{lem:isometry}), which follows from reversibility and inner-product invariance of the dynamics,
\begin{equation}
    \bigl\| \Delta W_{AB}(\tau) \bigr\|
    = \bigl\| \Delta W_{AB}(0) \bigr\|.
\end{equation}

\textit{Step 3: Marginalization cannot increase distinguishability.}
The harvester's final-state deviation is the marginal of the joint deviation over the source degrees of freedom:
\begin{equation}
    \Delta W_A(\tau)
    = \mathcal{M}_B\bigl[ \Delta W_{AB}(\tau) \bigr],
\end{equation}
where $\mathcal{M}_B$ is the unique marginalization map characterized in Proposition~\ref{proposition_uniqueness_marginalization}. By the contractivity of marginalization with respect to the base norm ( Lemma~\ref{lem:marginal_contractivity}), we have
\begin{equation}
    \bigl\| \Delta W_A(\tau) \bigr\|
    = \bigl\| \mathcal{M}_B[ \Delta W_{AB}(\tau) ] \bigr\|
    \le \bigl\| \Delta W_{AB}(\tau) \bigr\|.
\end{equation}

\textit{Step 4: Chaining the bounds.}
Combining the relations obtained in Steps~1–3, we find
\begin{align}
    \bigl\| \Delta W_A(\tau) \bigr\|
    &\le \bigl\| \Delta W_{AB}(\tau) \bigr\|
        &&\text{(contractivity of marginalization, Step 3)} \nonumber\\
    &=    \bigl\| \Delta W_{AB}(0) \bigr\|
        &&\text{(isometry under reversible evolution, Step 2)} \nonumber\\
    &=    \bigl\| \Delta W_B(0) \bigr\|
        &&\text{(product composition of the base norm, Step 1)}.
\end{align}
Thus
\begin{equation}
    \bigl\| \Delta W_A(\tau) \bigr\|
    \;\le\;
    \bigl\| \Delta W_B(0) \bigr\|,
\end{equation}
which completes the proof.
\end{proof}

\subsection{Classical versus quantum implications of the distinguishability bound}
\label{sec:classical_quantum_implications}

In this subsection we make precise the statement in the main text that the distinguishability bound (bound (5) in the main text)
\begin{equation}
D\bigl(W_{A,1}(\tau),W_{A,2}(\tau)\bigr)
\;\le\;
D\bigl(W_{B,1},W_{B,2}\bigr)
\label{eq:SM_D_bound}
\end{equation}
has very different implications in classical and quantum regimes.
The key point is that the same mathematical inequality can be physically trivial for idealized classical point-particle states, but nontrivial for quantum states forced by the uncertainty principle to occupy a finite phase-space volume.

\subsubsection{Classical point-like pure states: discontinuity of $D$.}

In classical mechanics, a pure state of a single particle is typically idealized as a point in phase space, represented by a Dirac delta distribution
\begin{equation}
 W_{\mathrm{cl}}(\bm{z}) = \delta(\bm{z} - \bm{z}_0),
 \qquad \bm{z} = (q,p).
\end{equation}
The base norm defined in Eq.~\eqref{eq:op_norm_def} reduces in this case to the $L_1$ (total variation) distance between probability densities.
Consider two such classical pure states located at different phase-space points,
\begin{equation}
 W_{\mathrm{cl},1}(\bm{z}) = \delta(\bm{z}-\bm{z}_1),
 \qquad
 W_{\mathrm{cl},2}(\bm{z}) = \delta(\bm{z}-\bm{z}_2),
 \qquad \bm{z}_1 \neq \bm{z}_2.
\end{equation}
Then their distance is
\begin{align}
 \bigl\|W_{\mathrm{cl},1} - W_{\mathrm{cl},2}\bigr\|
 &= \int d\bm{z}\, \bigl|\delta(\bm{z}-\bm{z}_1) - \delta(\bm{z}-\bm{z}_2)\bigr| \nonumber\\
 &= \int d\bm{z}\,\delta(\bm{z}-\bm{z}_1) + \int d\bm{z}\,\delta(\bm{z}-\bm{z}_2) \nonumber\\
 &= 1 + 1 = 2,
\end{align}
for any nonzero displacement $\bm{z}_2-\bm{z}_1$.
If $\bm{z}_2=\bm{z}_1$ the distance is trivially zero.
Hence we have a discontinuity
\begin{equation}
 \lim_{\bm{z}_2\to\bm{z}_1} \bigl\|W_{\mathrm{cl},1} - W_{\mathrm{cl},2}\bigr\| = 2 \neq 0,
\end{equation}
even though the phase-space displacement tends to zero.
Operationally, this reflects the fact that ideal classical point states are perfectly distinguishable: any infinitesimal parameter mismatch already yields the maximal distinguishability $D=2$.

As a consequence, the right-hand side of the bound~\eqref{eq:SM_D_bound} becomes saturated for any nonzero perturbation of a classical point-like source state.
The inequality is then mathematically true but physically trivial: it does not impose any restriction on how large $D(W_{A,1}(\tau),W_{A,2}(\tau))$ may become, and therefore does not rule out strong sensitivity to initial conditions (chaos) in the usual classical sense.

\subsubsection{Quantum minimum-uncertainty states: continuity of $D$.}

In quantum mechanics, by contrast, physically allowed pure states cannot be arbitrarily localized in phase space due to the uncertainty principle $\Delta q\,\Delta p \ge \hbar/2$.
A natural family of ``most classical'' pure states are the coherent states $|\alpha\rangle$, with
\begin{equation}
 \alpha = \frac{q+ip}{\sqrt{2\hbar}},
 \qquad \bm{z} = (q,p),
\end{equation}
so that different phase-space points correspond to different coherent states.

For two pure states $|\psi\rangle,|\phi\rangle$, the base norm (trace distance) between the corresponding density operators is
\begin{equation}
 \bigl\||\psi\rangle\langle\psi| - |\phi\rangle\langle\phi|\bigr\|
 = 2\sqrt{1 - |\langle\psi|\phi\rangle|^2}.
\end{equation}
Taking $|\psi\rangle = |\alpha\rangle$ and $|\phi\rangle = |\alpha+\delta\alpha\rangle$, we have
\begin{equation}
 |\langle\alpha|\alpha+\delta\alpha\rangle|^2
 = \exp\bigl(-|\delta\alpha|^2\bigr),
\end{equation}
and hence
\begin{align}
 D\bigl(W_\alpha, W_{\alpha+\delta\alpha}\bigr)
 &= \bigl\||\alpha\rangle\langle\alpha| - |\alpha+\delta\alpha\rangle\langle\alpha+\delta\alpha|\bigr\| \nonumber\\
 &= 2\sqrt{1 - e^{-|\delta\alpha|^2}}.
\end{align}
Using $|\delta\alpha| = |\delta\bm{z}|/\sqrt{2\hbar}$ and expanding for small displacements $|\delta\bm{z}|\ll\sqrt{\hbar}$, we obtain
\begin{align}
 D\bigl(W_\alpha, W_{\alpha+\delta\alpha}\bigr)
 &\approx 2\sqrt{1 - \Bigl(1 - \frac{|\delta\bm{z}|^2}{2\hbar}\Bigr)} \nonumber\\
 &= \sqrt{\frac{2}{\hbar}}\,|\delta\bm{z}|.
\end{align}
Thus, in the quantum case the distinguishability $D$ is a \emph{continuous} function of the physical displacement in phase space: as $|\delta\bm{z}|\to 0$, we have $D\to 0$ linearly.

\subsubsection{ Summary: same bound, different physical content.}

In both classical and quantum mechanics the generalized Hamiltonian evolution considered here acts as an isometry for $D$ on closed systems: it preserves the base norm distance between any pair of states.
Our distinguishability bound~\eqref{eq:SM_D_bound} then follows from this isometry together with the contractivity of discarding subsystem $B$.

The crucial difference between the two regimes is not in the dynamics, but in which initial states are physically allowed:

\begin{itemize}
 \item In classical mechanics, idealized point-like states (delta functions) are permitted and are mutually perfectly distinguishable for any nonzero displacement. For such states, $D$ is discontinuous in the phase-space parameters and the right-hand side of~\eqref{eq:SM_D_bound} is trivially saturated even for arbitrarily small perturbations. The bound therefore does not exclude classical sensitivity to initial conditions.

 \item In quantum theory, the uncertainty principle forbids such point-like states. All physically realizable pure states have finite phase-space extent, and $D$ between nearby states is continuous and small for small parameter displacements. In this case, a small perturbation of the source (small $D(W_{B,1},W_{B,2})$) necessarily implies a small distinguishability $D(W_{A,1}(\tau),W_{A,2}(\tau))$ at the harvester, so the bound~\eqref{eq:SM_D_bound} becomes a genuine robustness statement.
\end{itemize}

This makes precise in what sense quantum theory has a robustness advantage for deterministic energy harvesting: the same mathematical distinguishability bound is physically nontrivial in the quantum regime but essentially trivial for idealized classical point-particle states.

\section{Different decompositions of the same state}

\begin{figure*}[htbp!]
  \centering

  \begin{minipage}[t]{0.22\textwidth}
    \includegraphics[width=\textwidth]{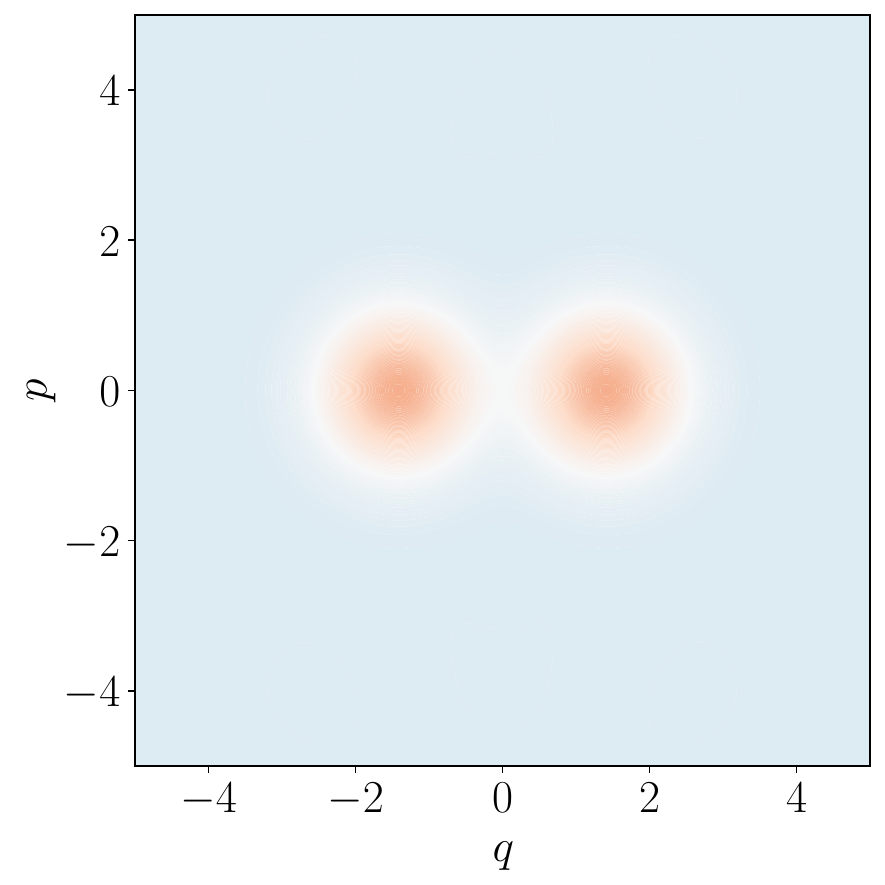}
    \centering (a)
  \end{minipage}
  \hfill
  \begin{minipage}[t]{0.22\textwidth}
    \includegraphics[width=\textwidth]{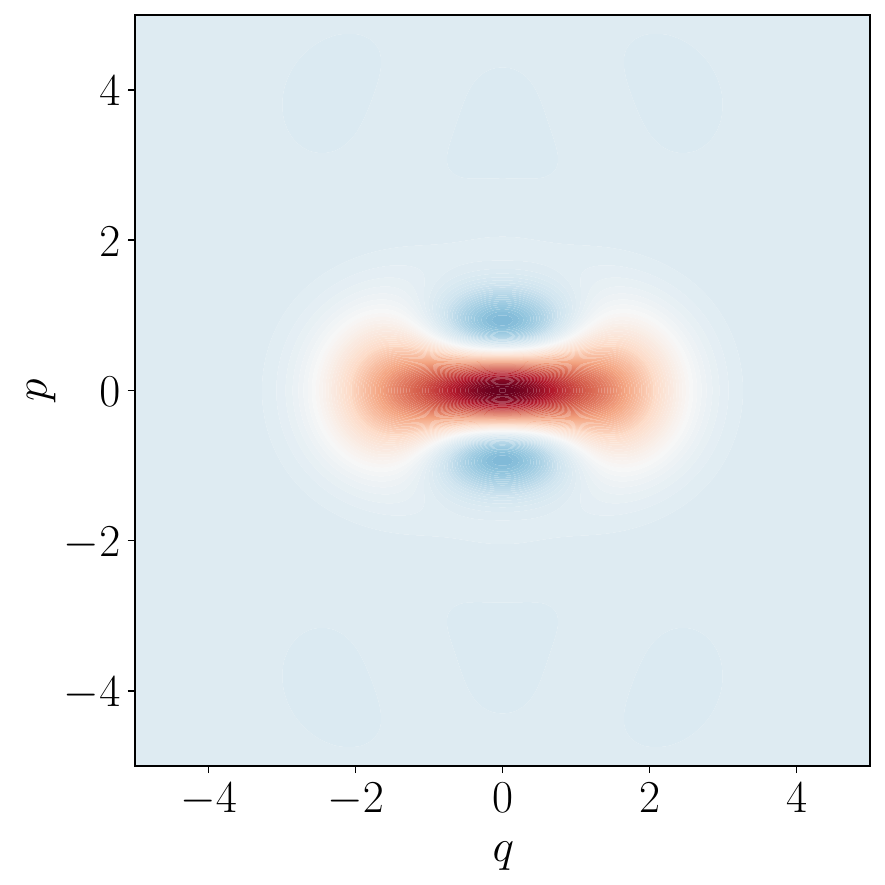}
    \centering (b)
  \end{minipage}
  \hfill
  \begin{minipage}[t]{0.22\textwidth}
    \includegraphics[width=\textwidth]{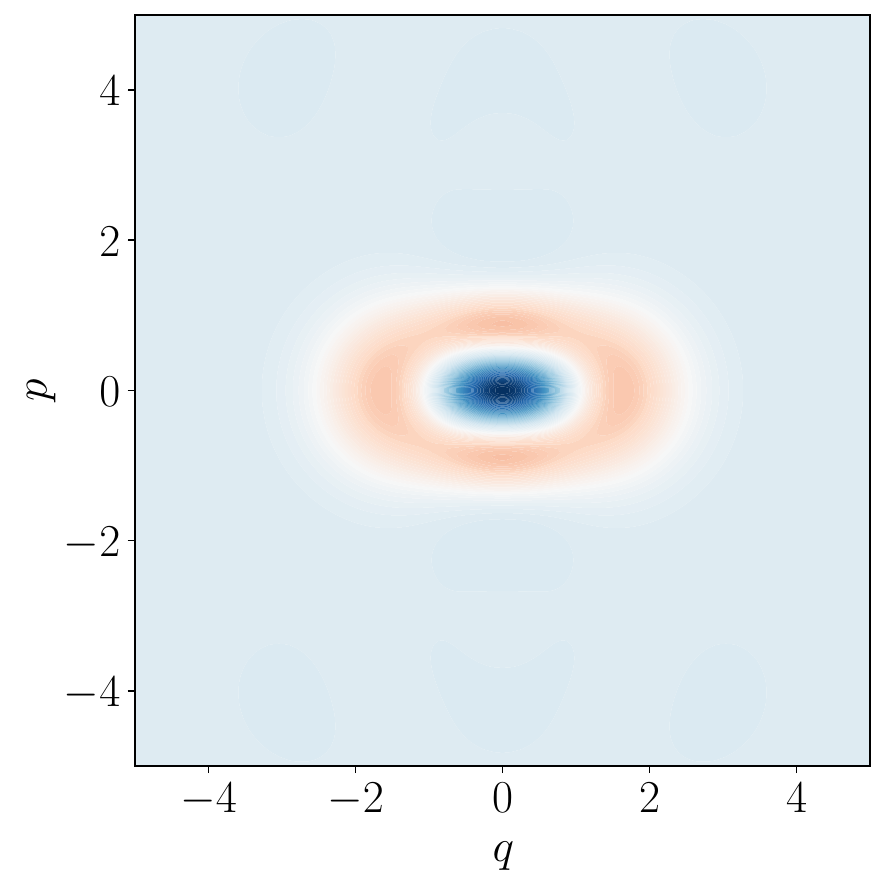}
    \centering (c)
  \end{minipage}
  \hfill
  \begin{minipage}[t]{0.22\textwidth}
    \includegraphics[width=\textwidth]{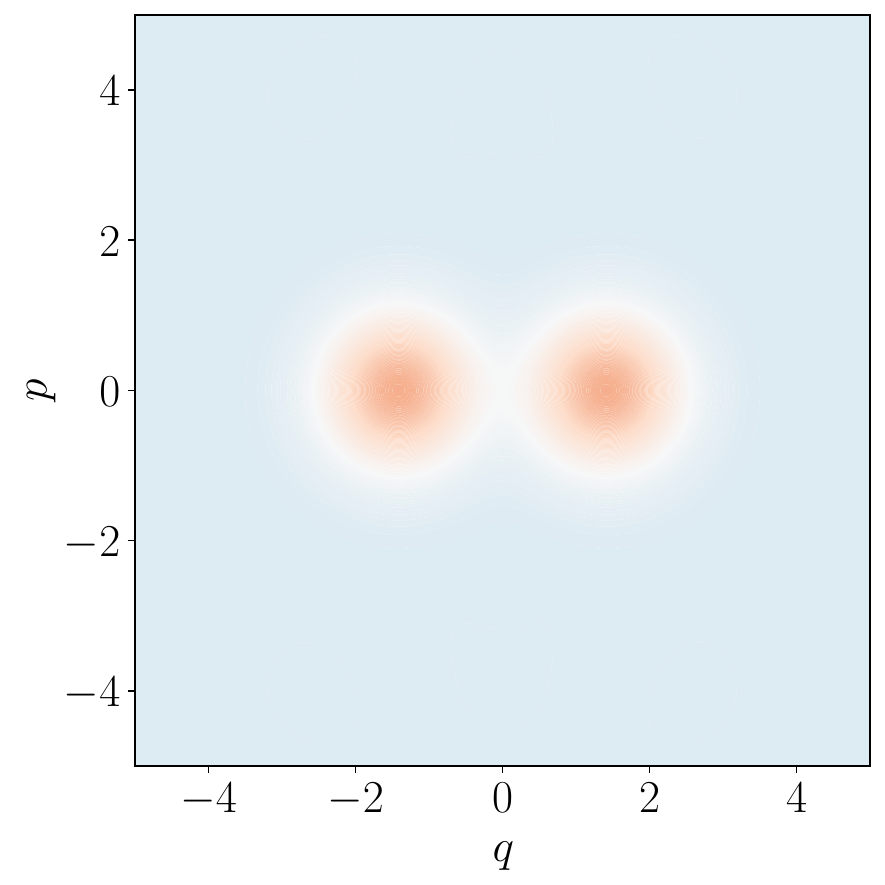}
    \centering (d)
  \end{minipage}
  \hfill
  \raisebox{1.8em}{ 
    \begin{minipage}[t]{0.05\textwidth}
      \includegraphics[width=\textwidth]{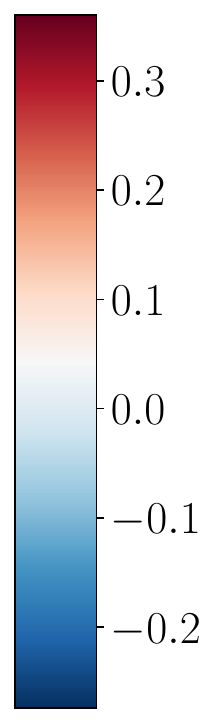}
    \end{minipage}
  }

  \captionsetup{justification=justified,singlelinecheck=false}
  \caption{\justifying
        Wigner function distributions for four quantum states of system B, evaluated at coherent state amplitude \(\alpha = 1\). Panel (a) displays the Wigner function of the incoherent mixture \(\rho_{B1} = \tfrac{1}{2}(|\alpha\rangle\langle\alpha| + |{-\alpha}\rangle\langle{-\alpha}|)\). Panel (b) corresponds to the even Schrödinger cat state \(\rho_{B2} = \tfrac{1}{2}|+\rangle_{\alpha}\langle +|_{\alpha}\), and panel (c) to the odd cat state \(\rho_{B3} = \tfrac{1}{2}  |-\rangle_{\alpha}\langle -|_{\alpha}\). Panel (d) shows \(\rho_{B4} = \tfrac{1}{2}(\rho_{B2} + \rho_{B3})\), a convex combination of even and odd cat states yielding the same density matrix as \(\rho_{B1}\) but with a different pure-state decomposition. The vertical colorbar represents the values of the Wigner function.
    }
    \label{fig:wigner_cats}
\end{figure*}

Even when the source field is in a mixed state, it could achieve DEH. Consider, for instance, the mixture
\begin{equation}
\rho_B = \frac{1}{2} \left( |\alpha\rangle\langle \alpha| + |{-\alpha}\rangle\langle {-\alpha}| \right),
\end{equation}
which admits a decomposition into the even and odd cat states defined as
\begin{equation}
|\pm \rangle_{\alpha} \approx \frac{1}{\sqrt{2}} (|\alpha\rangle \pm |{-\alpha}\rangle).
\end{equation}
These cat states have been extensively studied and experimentally realized in platforms such as cavity QED and superconducting circuits. Although \(\rho_B\) is already a convex combination of two coherent states, it can also be expressed as
\begin{equation}
\rho_B = \frac{1}{2} \left( |+\rangle_{\alpha}\langle +|_{\alpha} + |-\rangle_{\alpha}\langle -|_{\alpha} \right).
\end{equation}
The Wigner functions corresponding to these representations are shown in Fig.~\ref{fig:wigner_cats}: the mixture in panel~(a), the even and odd cat states in panels~(b) and~(c), respectively, and the alternative decomposition in panel~(d). This example illustrates the non-uniqueness of mixed-state decompositions.

Despite these distinct decompositions, evolution under \(H_{\text{JC}}\) yields the same reduced state \(\rho_A(t)\) for the qubit, demonstrating the convex closure property of DEH described in Proposition~\ref{prop:DEH_quantum_convexclosure}.

\section{Approximate DEH protocol: relative entropy argument in quantum case}

In the development of the approximate DEH protocol, a key requirement is that small deviations in the energy source should not lead to amplified errors in the energy harvester. Mathematically, if the initial states \(\rho_{B1}\) and \(\rho_{B2}\) satisfy \(S(\rho_{B1} \| \rho_{B2}) = \varepsilon\), then the corresponding output states \(\rho_{A1}(\tau)\) and \(\rho_{A2}(\tau)\), resulting from unitary evolution and partial trace stopping at the proper time \(t=\tau\), satisfy
\[
S(\rho_{A1}(\tau) \| \rho_{A2}(\tau)) \leq \varepsilon.
\]
This ensures that the relative entropy — a measure of distinguishability — does not increase during the protocol, providing robustness against imperfections in source preparation.

However, closeness between \(\rho_{A1}(\tau)\) and \(\rho_{A2}(\tau)\) alone does not guarantee effective energy harvesting. It remains possible that both states are close to each other, yet far from the desired excited state \(|1\rangle\langle 1|\), rendering the protocol ineffective. To resolve this, we assume that \(\rho_{A1}(\tau)\) is close to \(|1\rangle\langle 1|\) — quantified by \(S(|1\rangle\langle 1| \| \rho_{A1}) = \mu\) — and aim to show that \(\rho_{A2}(\tau)\) must also be close to \(|1\rangle\langle 1|\). This closes the gap in the approximate DEH analysis and ensures the protocol’s operational success. Starting from below, we will omit \((\tau)\) and use \(\rho_{A1}\) and \(\rho_{A2}\) to represent the final states of the qubit.

The quantum Pinsker inequality \cite{hiai1991proper} bounds the trace distance in terms of relative entropy,
\begin{equation}
\frac{1}{2} \lVert \rho - \sigma \rVert_1^2 \leq \ln 2 \; S(\rho \| \sigma),
\end{equation}
so that for the approximate DEH protocol,
\begin{equation}
\lVert \rho_{A1} - \rho_{A2} \rVert_1 \leq \sqrt{2 \ln 2 \, \varepsilon}, \qquad
\lVert |1\rangle\langle 1| - \rho_{A1} \rVert_1 \leq \sqrt{2 \ln 2 \, \mu}.
\end{equation}
Since the trace norm is a metric, the triangle inequality implies
\begin{equation}
\lVert |1\rangle\langle 1| - \rho_{A2} \rVert_1 
\leq \lVert |1\rangle\langle 1| - \rho_{A1} \rVert_1 + \lVert \rho_{A1} - \rho_{A2} \rVert_1
=: \delta,
\end{equation}
with \(\delta = \sqrt{2 \ln 2 \, \mu} + \sqrt{2 \ln 2 \, \varepsilon}\).

To bound the relative entropy of \(\rho_{A2}\) with respect to the ideal excited state, we employ Audenaert’s dimension-independent inequality \cite{audenaert2005continuity},
\begin{equation}
S(\rho \| \sigma) \leq T \log d + \min(-T \log T, 1/e) - \frac{T \log \lambda_{\min}(\sigma)}{2},
\end{equation}
where \(T = \lVert \rho - \sigma \rVert_1\), \(d\) is the Hilbert space dimension, and \(\lambda_{\min}(\sigma)\) the smallest eigenvalue of \(\sigma\). Setting \(\rho = |1\rangle\langle 1|\), \(\sigma = \rho_{A2}\), \(d=2\), and \(T = \delta\), we obtain
\begin{equation}
S(|1\rangle\langle 1| \| \rho_{A2}) \leq 
\delta \log 2 + \min(-\delta \log \delta, 1/e) - \frac{\delta \log \lambda_{\min}(\rho_{A2})}{2}.
\end{equation}

Thus, if \(\rho_{A1}\) is close to \(|1\rangle\langle 1|\) and \(\rho_{A2}\) is close to \(\rho_{A1}\) in trace distance, it follows that \(\rho_{A2}\) is also close to \(|1\rangle\langle 1|\) in relative entropy. This establishes the approximate DEH protocol’s robustness against small deviations in source preparation.

\section{Entropy cycle of the qubit}

An intriguing feature of the quantum energy harvesting protocol lies in the dynamics of entropy transfer between the energy source and the harvesting system. When the energy source is initially prepared in a mixed state, its von Neumann entropy is strictly nonzero. However, our analysis reveals that, throughout the harvesting process, this entropy is not transferred to the harvesting system. Despite energy exchange between the two systems, the entropy of the harvesting system remains nearly unchanged when the energy harvesting process is stopped at a proper time.

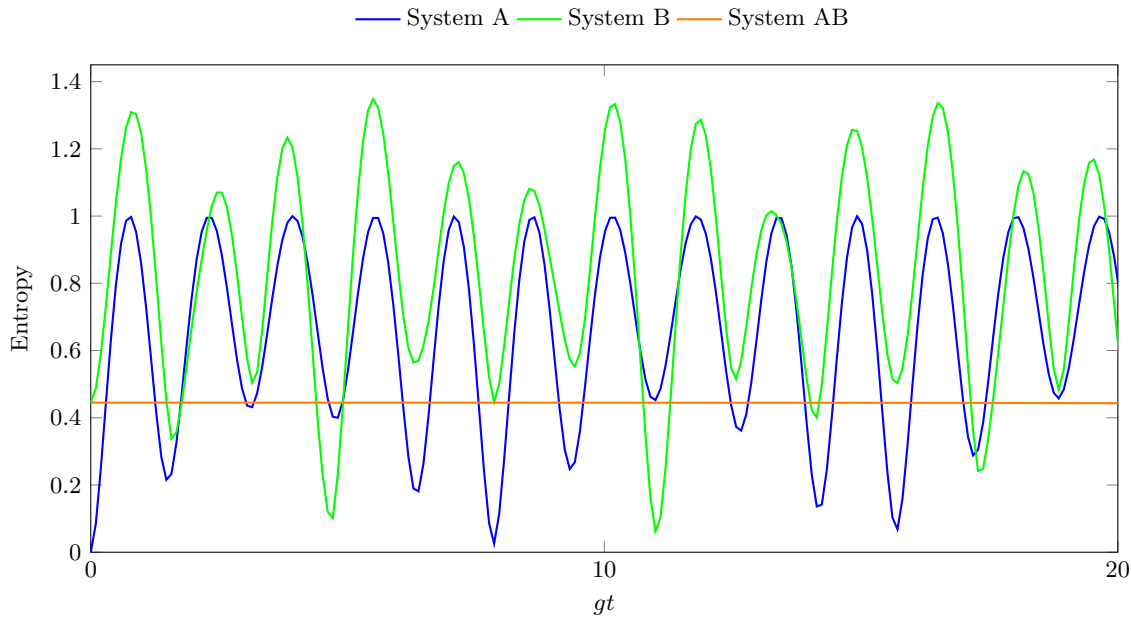
\begin{figure}[htbp!]
    \centering
    \begin{subfigure}{\linewidth}
        \centering
        \begin{tikzpicture}
\begin{axis}[
    width=0.85\linewidth,
    height=0.45\linewidth,
    xlabel={\(g t\)},
    ylabel={Entropy},
    xmin=0, xmax=20,
    ymin=0, ymax=1.45,
    xtick distance=10,
    ytick distance=0.2,
    legend style={draw=none, at={(0.5,1.05)}, anchor=south, font=\small, legend columns=-1},
    tick label style={font=\small},
    label style={font=\small},
    domain=0:25,
    samples=2000
]

\addplot [blue, thin, thick] table [x=t, y=entropy_A, col sep=space] {entropy_data_1.dat};
\addlegendentry{System A}

\addplot [green, thin, thick] table [x=t, y=entropy_B, col sep=space] {entropy_data_1.dat};
\addlegendentry{System B}

\addplot [orange, thin, thick] table [x=t, y=entropy_AB, col sep=space] {entropy_data_1.dat};
\addlegendentry{System AB}

\end{axis}
\end{tikzpicture}
        \label{fig:DEH_entropy_sub}
    \end{subfigure}

    \caption{\justifying
    Entropy dynamics of each subsystem during the energy harvesting process governed by the Jaynes--Cummings interaction. The plot shows that, by stopping the interaction at an appropriate time, one can achieve energy transfer with almost no net entropy transfer, highlighting the reversibility of the process under optimal timing conditions.}
    \label{fig:DEH_entropy}
\end{figure}

Figure~\ref{fig:DEH_entropy} displays such entropy evolution of System A (blue), System B (green), and the joint system AB (orange) during the quantum energy harvesting process governed by the Jaynes–Cummings interaction. In this example, the energy source (System B) is initialized in a mixed state with amplitude \(\vert \alpha \vert^2=0.1\), introducing nonzero entropy into the dynamics. Both subsystems A and B exhibit coherent, quasi-periodic oscillations in their von Neumann entropy, reflecting the reversible exchange of quantum information driven by the interaction. In contrast, the entropy of the composite system remains nearly constant throughout, as expected for unitary evolution. The oscillatory behavior of the individual subsystems indicates that while energy is being exchanged, entropy is dynamically redistributed without net increase. Crucially, the entropy of the harvesting system returns close to its initial value at specific times, demonstrating that energy extraction can occur with negligible entropy gain. These results exemplify the central feature of the protocol: it enables energy transfer without accompanying entropy transfer, preserving the purity of the harvester and realizing a thermodynamically efficient mechanism in the quantum regime.

Moreover, we investigate how different decompositions of the same initial mixed state affect the entropy dynamics. The time evolution of the system reveals a consistency: different decompositions yield identical oscillatory patterns in the entropy of the harvesting system. This invariance suggests that the entropy oscillations are determined solely by the mixed state's density matrix and are insensitive to the specific decomposition preparations. The entropy cycle, therefore, appears to be a robust feature of the protocol.

\section{Compatibility of the fully quantized and the semiclassical Hamiltonians - an example of the DEH model}

By using Liouville-Von Neumann equation, we can get the time evolution of the density matrix like this:

\begin{equation}\label{A1}
\frac{\partial \rho_{AB}}{\partial t} = \frac{i}{\hbar} \ [\rho_{AB},\ H].
\end{equation}

In order to get the effective Hamiltonian on system A, we need to take the partial trace on system B. Therefore, we get the following derivation. Noticing that We are using a similar Hamiltonian to the Jaynes-Cummings model (JCM). However, the difference is that we do not use rotating wave approximation (RWA) to ignore the fast rotating terms here, so a more general Hamiltonian is applied in the derivation:

\begin{align}\label{A2}
\frac{\partial \rho_{A}}{\partial t} 
&=tr_B \{{\frac{\partial \rho_{AB}} {\partial t}}\} \nonumber \\
&=tr_B \{\frac{i}{\hbar} \ [\rho_{AB},\ H]\} \nonumber \\
&=\frac{i}{\hbar}\ tr_B \{[\rho_{A} \otimes |\alpha\rangle\langle \alpha|, \ H]\} \nonumber \\
&=\frac{i}{\hbar}\ tr_B \{[\rho_{A} \otimes |\alpha\rangle\langle \alpha|, \ \frac{\hbar \omega_0}{2} (\hat{\sigma}_z \otimes I) + \hbar \omega_c (I\otimes \hat{a}^\dagger \hat{a}) + \hbar g (\hat{\sigma}_x \otimes (\hat{a} +  \hat{a}^\dagger))]\} \nonumber \\
&=\frac{i}{\hbar}\ tr_B \{\frac{\hbar \omega_0}{2}(\rho_{A} \hat{\sigma}_z \otimes |\alpha\rangle\langle \alpha|- \hat{\sigma}_z \rho_{A}\otimes |\alpha\rangle\langle \alpha|) \nonumber \\
&{\hspace{32pt}}+ \hbar \omega_c (\rho_{A} \otimes |\alpha\rangle\langle \alpha|\ \hat{a}^\dagger \hat{a} - \rho_{A} \otimes \hat{a}^\dagger \hat{a} \  |\alpha\rangle\langle \alpha|) \nonumber \\
&{\hspace{32pt}}+ \hbar g (\rho_{A}\hat{\sigma}_x \otimes |\alpha\rangle\langle \alpha|(\hat{a} +\hat{a}^\dagger) - \hat{\sigma}_x\rho_{A} \otimes (\hat{a} +\hat{a}^\dagger)|\alpha\rangle\langle \alpha|)\} \nonumber \nonumber \\
&=\frac{i}{\hbar} \frac{1}{2}\hbar\omega_0\ [\rho_{A},\ \hat{\sigma}_z] + \frac{i}{\hbar} \hbar g[\rho_{A}, \hat{\sigma}_x]\ tr_B\{|\alpha\rangle\langle \alpha|\ (\hat{a} +\hat{a}^\dagger)\} \nonumber \\
&=\frac{i}{\hbar} \frac{1}{2}\hbar\omega_0\ [\rho_{A},\ \hat{\sigma}_z] + \frac{i}{\hbar} \sqrt{2}\hbar g[\rho_{A}, \hat{\sigma}_x]\langle\hat{x}\rangle_{\rho_B}.
\end{align}

Hence, we can get the effective Hamiltonian on system A:

\begin{equation}\label{A3}
H^A_{eff} = \frac{\hbar\omega_0}{2}\hat{\sigma}_z + \sqrt{2}\hbar g\hat{\sigma}_x\langle\hat{x}\rangle_{\rho_B}
\end{equation}

To align with the Hamiltonian presented in \cite{meng2025quantum},we focus on the variable $\langle\hat{x}\rangle_{\rho_B}$ as follows:

\begin{equation}\label{A4}
\langle \hat{x} \rangle_{\rho_B} 
= \langle \alpha | \hat{x} | \alpha \rangle 
= \langle 0 | D^\dagger(\alpha) \hat{x} D(\alpha) | 0 \rangle 
= \langle 0 | \hat{x} + \frac{\alpha + \alpha^*}{\sqrt{2}} | 0 \rangle 
= \frac{\alpha + \alpha^*}{\sqrt{2}}.
\end{equation}

Next, we transform back to the interaction picture, where $e^{-iHt/\hbar}|\alpha\rangle=|\alpha(t)\rangle$ and $\alpha(t)=e^{-i\omega t}\alpha$. Additionally, we express $\alpha$ as $|\alpha|e^{-i\phi}$. Thus, equation \eqref{A4} becomes:

\begin{equation}\label{A5}
\langle \hat{x} \rangle_{\rho_B}(t) 
= \frac{\alpha(t) + \alpha^*(t)}{\sqrt{2}} 
= \frac{1}{\sqrt{2}} \big( e^{-i\omega t} |\alpha| e^{-i\phi} + e^{i\omega t} |\alpha| e^{i\phi} \big) 
= \frac{|\alpha|}{\sqrt{2}} \big( e^{-i(\omega t+\phi)} + e^{i(\omega t+\phi)} \big) 
= \sqrt{2} |\alpha| \cos(\omega t + \phi).
\end{equation}

\noindent and in the end, the effective Hamiltonian becomes this with the assumption that $\hbar=1$ and $A= g|\alpha|$:

\begin{equation}\label{A6}
H^A_{eff}=\frac{E}{2}\hat{\sigma}_z+2A\cos(\omega t+\phi)\hat{\sigma}_x,
\end{equation}

\noindent which aligns with the semiclassical model. If we Use the JCM Hamiltonian, which incorporates the RWA in the above computation, one can also obtain another version of the semiclassical model that results from applying the RWA, as in the appendix of \cite{meng2025quantum}.

The question is under what conditions one can compute an "effective Hamiltonian" for system~A. In general, the full Hamiltonian includes interactions between systems~A and~B, so one cannot simply neglect the back-action of B when tracing it out. However, in the semiclassical limit, such an approximation becomes valid, as discussed in \cite{twyeffort2022defining}. 

In the limit where the field is weakly excited ($|\alpha| \ll 1$) or in the semiclassical regime ($g \to 0$, $|\alpha| \to \infty$), the characteristic collapse and revival pattern of the JCM is suppressed, and the system dynamics reduce to simple Rabi oscillations. In particular, for a classical light field that is much stronger than the quantized source ($|\alpha| \to \infty$), the influence of the harvester on the source can be neglected. This corresponds to the standard assumption in the semiclassical model that backaction from system~A to system~B is negligible. Physically, when the source contains a large number of photons, on the order of $10^{10}$--$10^{15}$, the reduced state of the field $\rho_B$ remains essentially unchanged during the energy-harvesting process.

Within this approximation, the fully quantized Hamiltonian in equation \eqref{A2} reduces to the semiclassical form \eqref{A6} upon identifying $A = g |\alpha|$, justifying the use of the effective Hamiltonian for system~A. Analytical calculations and numerical simulations confirm that in this limit the dynamics of the fully quantized DEH model and the semiclassical model are in excellent agreement, demonstrating their compatibility. Conversely, at low photon numbers, backaction from the harvester becomes significant, which changes the source state, and the semiclassical approximation fails to capture the full quantum dynamics, leading to observable deviations from the JCM evolution.





\end{document}